\documentclass[pra,twocolumn,superscriptaddress,showpacs,amsmath,amstex,amssymb,citeautoscript]{revtex4-1}

\newcommand{\vect}[1]{\boldsymbol{#1}}
\newcommand{\pdag}[0]{{\phantom{\dagger}}}

\usepackage[T1]{fontenc}
\usepackage[utf8]{inputenc}
\usepackage{lipsum}
\usepackage{amsmath}
\usepackage{amssymb}
\usepackage{bbm}
\usepackage{braket}
\usepackage{xcolor}
\usepackage{pifont}
\usepackage[mathscr]{euscript}
\usepackage[shortlabels]{enumitem}
\usepackage[justification=justified]{subcaption}
\usepackage{graphicx}
\usepackage{lipsum}
\allowdisplaybreaks
\usepackage{float}
\usepackage{graphicx}
\usepackage{dsfont}
\usepackage{comment}
\usepackage[colorlinks=true]{hyperref}  
\hypersetup{
    bookmarks=true,         
    unicode=false,          
    pdftoolbar=true,        
    pdfmenubar=true,        
    pdffitwindow=false,     
    pdfstartview={FitH},    
    pdftitle={Band-Off-Diagonal Pairing},    
    pdfauthor={Putzer and Scheurer},     
    pdfsubject={},   
    pdfcreator={},   
    pdfproducer={}, 
    pdfkeywords={} {} {}, 
    pdfnewwindow=true,      
    colorlinks=true,       
    linkcolor=blue, 
    citecolor=blue,        
    filecolor=magenta,      
    urlcolor=blue           
} 

\usepackage{cleveref}
\crefname{figure}{Fig.}{Figs.}

\newcommand{\equref}[1]{Eq.~(\ref{#1})}

\newcommand{\secref}[1]{Sec.~\ref{#1}}
\newcommand{\figref}[1]{Fig.~\ref{#1}}
\newcommand{\refcite}[1]{Ref.~\cite{#1}}

\newcommand{\tableref}[1]{Table~\ref{#1}}
\newcommand{\appref}[1]{Appendix~\ref{#1}}

\newcommand{\sign}{\,\text{sign}}
\renewcommand{\approx}{\simeq}

\newcommand{\ie}{i.e.~}

\renewcommand{\vec}[1]{\boldsymbol{#1}}
\definecolor{wrongultramarine}{rgb}{1,0.5,0}

\newcommand{\vk}{{\boldsymbol{k}}}
\newcommand{\vq}{{\boldsymbol{q}}}
\newcommand{\bdg}{\mathrm{BdG}}
\newcommand{\imdt}{{\Delta_2^I}}
\newcommand{\conv}[1]{\mathrm{conv_{#1}}}
\newcommand{\geo}[1]{\mathrm{geo_{#1}}}

\begin{document}

\title{Eliashberg Theory and Superfluid Stiffness \\ of Band-Off-Diagonal Pairing in Twisted Graphene}

\author{Bernhard Putzer}
\affiliation{Institute for Theoretical Physics, University of Innsbruck, Innsbruck A-6020, Austria}
\affiliation{Institute for Theoretical Physics III, University of Stuttgart, 70550 Stuttgart, Germany}

\author{Mathias S.~Scheurer}
\affiliation{Institute for Theoretical Physics III, University of Stuttgart, 70550 Stuttgart, Germany}

\begin{abstract}
Recently, band-off-diagonal superconductivity has been proposed [\href{https://www.nature.com/articles/s41467-023-42471-4}{Nat.~Commun.~\textbf{14}, 7134 (2023)}] as a candidate pairing state for twisted graphene systems. Based on mean-field theory, it was shown that it not only naturally emerges from both intervalley electron-phonon coupling and fluctuations of the nearby correlated insulator, but also exhibits nodal and gapped regimes as indicated by scanning tunneling microscopy experiments. Here we study band-off-diagonal pairing within Eliashberg theory. We show that despite the additional frequency dependence, the leading-order description of both intervalley coherent fluctuations or intervalley phonons exhibits a symmetry prohibiting admixture of an intraband component to the interband pairing state. It is found that even- and odd-frequency pairing mix, which originates from the reduced number of flavor degrees of freedom in the normal state. From analytic continuation, we obtain the electronic spectral function showing that, also within Eliashberg theory, the interband nature leads to an enhanced spectral weight below the order-parameter energy compared to band-diagonal pairing. Finally, we also study the superfluid stiffness of band-off-diagonal pairing, taking into account multi-band and quantum geometry effects. It is shown that for $s$-wave and chiral momentum dependencies, conventionally leading to fully gapped phases, an interband structure reduces the temperature scale below which the stiffness saturates. Depending on parameters, for the chiral state, this scale can even be suppressed all the way to zero temperature leading to a complex competition of multiple dispersive and geometrical contributions. Our results show that interband pairing might also be able to explain more recent stiffness measurements in the superconducting state of twisted multilayer graphene. 
\end{abstract}

\maketitle

\section{Introduction}
As a result of the rich physics displayed by twisted layers of graphene \cite{review1,review2,review3}, there have been enormous research activities studying their properties in the last few years. Although experiments have been able to clarify certain aspects of their phase diagram, superconductivity, both concerning the form and symmetry of the order parameter and the underlying pairing mechanism, is still under debate. In this regard, local tunneling conductance measurements \cite{Oh_2021,TunnelingPerge} have provided strong constraints by revealing a V-shaped tunnel density of states, which---depending on filling and samples---can become more U-shaped. Apart from fluctuations as the origin \cite{vestigialSC}, this points to nodal to fully gapped transitions inside the superconducting phase. Taking into account the flavor polarization in the normal state in the relevant filling fraction between $\nu=2$ and $3$ \cite{Wong_2020,Zondiner_2020,Park_2021,Hao_2021,DiodeEffect,morissetteElectronSpinResonance2022,PauliLimit,PhysRevB.98.054515,OurClassification,PhysRevB.106.104506,Christos_2020,Khalaf_2021,2022PhRvX..12b1018C,PhysRevB.105.224508,Scammell_2022,PhysRevLett.127.247703,PhysRevB.105.094506,2024arXiv240409909Z,2023arXiv230800748I}, \refcite{BandOffDiagonalPairing} proposed band-off-diagonal (BOD) pairing as a possibility to naturally capture both gapped and nodal regimes. To describe this form of pairing, it is important to keep both quasi-flat low-energy bands (index $\alpha$) in each valley (for the respective active spin flavor). The superconducting order parameter, which is now a matrix $\Delta_{\alpha,\alpha’}(\vec{k})$ in band space, for the BOD pairing state has the form $\Delta(\vec{k}) = \chi_{\vec{k}}\sigma_y$ where $\chi_{\vec{k}} \in \mathbb{C}$ captures the momentum ($\vec{k}$) dependence. It couples, as usual, electronic states in opposite valleys and of opposite momenta but in distinct bands; since these bands are not related by time-reversal symmetry, they are not exactly degenerate, which crucially alters the spectral function. This can lead to transitions from nodal to fully gapped pairing states as a function of system parameters such as filling, pairing strength, and band splitting.

To understand why such a state can be favored energetically, consider the scenario where the dominant interaction is attractive in the intervalley Cooper channel---this will, in fact, be the case if phonons \cite{PhysRevLett.121.257001,PhysRevLett.122.257002,Lewandowski2021,PhononsAndBandren,2022arXiv220202353Y,PhysRevB.104.L121116} coupling the two valleys, which have recently been experimentally identified as dominant \cite{2023arXiv230314903C,2023arXiv230315551L}, provide the pairing glue or if fluctuations of the nearby correlated insulator, a time-reversal symmetric intervalley coherent phase \cite{TIVCExp}, stabilizes the superconductor. In that case, the leading order parameter will have the same sign in both valleys (at opposite momenta) and, by virtue of being totally antisymmetric, must obey $\Delta (\vec{k}) = -\Delta^T (\vec{k})$; this immediately leads to the form stated above.
We emphasize that, as discussed in \refcite{BandOffDiagonalPairing}, a purely BOD pairing state is still well defined if there is additional T-IVC order in the normal state \cite{TIVCExp}. In that case, the two bands $\alpha=\pm$ refer to admixtures of the two valleys. Also, as before, the superconductor will be stabilized by intervalley phonons and can be fully gapped or nodal depending on parameters. The main difference from the case only with spin polarization and no intervalley coherence in the normal state is that, now, all other intra-band order parameters necessarily have nodal lines, as a result of Fermi-Dirac statistics.

In this work, we generalize the mean-field analysis in \refcite{BandOffDiagonalPairing} of BOD pairing to the Eliashberg framework, where the order parameter $\Delta_{\alpha,\alpha’}(\vec{k})$ is promoted to $\phi_{\alpha,\alpha’}(i\omega_n,\vec{k})$, which also depends on Matsubara frequency $\omega_n$. Recalling the abovementioned mechanism for BOD pairing, it is natural to ask whether the $\omega_n$ dependence might change the mean-field findings noticeably: the analogous property for attractive intervalley Cooper-channel interactions reads as $\phi (i\omega_n,\vec{k}) = -\phi^T (-i\omega_n,\vec{k})$ and now not only allows for BOD pairing but also for the admixture of a band-diagonal (odd-frequency) component. We show that there is an additional symmetry---exact in the limit where the bosonic modes mediating the interactions couple only with the leading, momentum-independent term to the fermions---that prohibits mixing of band-diagonal and off-diagonal components in $\phi$. In agreement with the mean-field study of \refcite{BandOffDiagonalPairing}, we thus find \textit{purely} BOD pairing for these types of interactions. Interestingly, we also reveal that the reduced flavor degrees of freedom do allow for admixture of even ($\propto \sigma_y$) and odd ($\propto \sigma_x$) frequency pairing; the admixed odd-frequency component increases with the splitting $t$ of the flat bands and vanishes when we take them to be degenerate, $t=0$. In line with \refcite{BandOffDiagonalPairing}, the key property of the BOD pairing state, setting it apart from the typical (mainly) band-diagonal superconductor, is that the gap energy scale is significantly reduced compared to the order parameter. We find the same within Eliashberg theory: upon performing a numerical analytic continuation to the real axis, the resultant spectral function and density of states of an approximately $\vec{k}$-independent order parameter looks significantly more V-shaped than that of a band-diagonal order parameter of the same strength; this confirms that BOD pairing can provide an explanation of the tunneling data \cite{Oh_2021,TunnelingPerge}.

In another more recent experiment \cite{banerjee_superfluid_2024}, the temperature dependence of the superfluid stiffness $D_s(T)$ \cite{EinzelPaper,Prozorov_2006,PhysRevX.9.031049, 2024arXiv240613740T, PhysRevLett.124.167002, PhysRevLett.123.237002} was probed and found to lack the exponential saturation behavior setting in below around $30\%$ of $T_c$ (see, e.g., \refcite{Prozorov_2006}) in typical fully gapped superconductors. We derive an expression for the superfluid stiffness for a BOD pairing state that also takes into account the quantum geometry of the Bloch states \cite{reviewQuantumGeom}. We find that, even for a $\vec{k}$-independent order parameter, the BOD nature can reduce the saturation temperature significantly and induce a non-monotonic behavior of  $\partial_T D_s(T)$, reminiscent of experiment \cite{banerjee_superfluid_2024}. A chiral BOD order parameter, which would generically be gapped in the band-diagonal scenario, can be nodal or fully gapped, depending on parameters \cite{BandOffDiagonalPairing}. We show that, in that case, the exponential saturation temperature of $D_s(T)$ can be suppressed all the way to the lowest temperature. We present a detailed low-$T$ analysis revealing a complex competition of multiple terms.

The remainder of the manuscript is organized as follows. In \secref{formalism}, we define the model we are studying, the Eliashberg framework, and study the symmetries of the superconducting gap equation. The numerical results, including the electronic spectral function, are presented and discussed in \secref{numericalresults}. Section~\ref{SuperfluidStiffness} is about the superfluid stiffness of BOD superconductors. Finally, the findings are summarized in \secref{Conclusion} and the appendices contain more details on the Gor'kov Green's function (\appref{app:nambu}), a discussion of other, more exotic, multi-band superconducting states (\appref{sec:broken_chiral_sym}), and more details on the superfluid stiffness (Appendix \ref{app:stiffness_correction} and \ref{sec:app_low_temp_scaling}).

\section{Multiband Eliashberg formalism}\label{formalism}

\subsection{General Theory}\label{sec:general_theroy}
We start by defining the interacting model for superconductivity that we study in this work. In the range of flat-band fillings $\nu$ with $2<|\nu| <3$, we assume that the normal state is already flavor-polarized. While additional coexistence with intervalley coherent order \cite{TIVCExp} might be realized at least in some samples and/or doping regimes, we here focus for simplicity on a spin-polarized normal state \cite{Microwave,PhysRevResearch.2.033062,PhysRevB.106.104506} and refer to \cite{BandOffDiagonalPairing} for a discussion of BOD pairing in the presence of intervalley coherence. With spin polarization, the effective low-energy model can be formulated using spinless electrons, with creation operators $c^\dagger_{\vk, \eta, \alpha}$ in one of the two valleys $\eta = \pm$ (coming from the valleys of the single-layer graphene), in one of the two bands $\alpha=\pm$ (related to the two bands forming the Dirac cones of the graphene layers), and at momentum $\vk$ in the 2D moiré Brillouin zone (MBZ).
We consider scattering between all these electronic degrees of freedom mediated by bosonic modes with creation operators $b^\dagger_{\vq,l}$ for branch $l$ and of momentum $\vq\in\mathrm{MBZ}$. As we will detail below, these bosons will primarily refer to optical phonon modes but the model can also be viewed as some effective description of an unconventional pairing mechanism where superconductivity is mediated by fluctuations of an order parameter of a correlated insulator~\cite{BandOffDiagonalPairing}. This leads to the Hamiltonian
\begin{subequations}\begin{equation}
    H=H_\mathrm{e} + H_\mathrm{b}+ H_\mathrm{e-b}
\end{equation}
where 
\begin{align}
     H_\mathrm{e} &=\sum_{\vk, \eta, \alpha}\xi_{\vk, \eta,\alpha} c^\dagger_{\vk, \eta,\alpha} c^\pdag_{\vk, \eta,\alpha} \label{DispersionTermxi}\\
    H_\mathrm{b} &= \sum_{\vq,l} \omega^l_\vq b^\dagger_{\vq,l} b^\pdag_{\vq,l} 
\end{align}
account for the bare electronic and bosonic energetics, with band energies $\xi_{\vk, \eta,\alpha}$ and $\omega^l_\vq$, respectively. For now, we will keep the coupling between the electrons and bosons general and parametrize it with the form factors $g_{\vk, \vk'}^l$ as follows: 
\begin{align}
    H_\mathrm{e-b} = \sum_{{\vk,\vk'},l,\alpha, \alpha', \eta,\eta'}
    &c^\dagger_{\vk, \eta,\alpha}\left(g_{\vk,\vk'}^l\right)_{\alpha, \alpha'}^{\eta,\eta'}c^\pdag_{\vk', \eta',\alpha'} \label{ElPhonCoupl} \\ \nonumber
    &\times\left[b^\pdag_{{\vk-\vk'},l} + b^\dagger_{{\vk'-\vk},l}\right]. 
\end{align}\end{subequations}
To respect the symmetries of alternating-twist-angle multilayer graphene, we consider $D_6$ as the system's point group, which will be further reduced to $C_6$ by application of a displacement field. 
The group elements of $D_6$, together with spinless time-reversal ($\Theta$), valley U(1) symmetry [$U(1)_v$], and chiral symmetry ($C$), and how they act on the electronic field operators and on the couplings $g_{\vk, \vk'}^l$ are given in \tableref{SymmetriesAndReps}.

\begin{table}[tb]
\begin{center}
\caption{Action of the symmetry-group elements $g$ on the fermionic operators and on the coupling matrix elements in \equref{ElPhonCoupl}. The last column shows what the respective symmetries imply for the electronic dispersion in \equref{DispersionTermxi}. $C_{n\hat{n}}$ denotes $n$-fold rotation along the $\hat{n}$ axis. Except for the anti-unitary time-reversal symmetry, $\Theta$, all representations are unitary; as usual, chiral symmetry refers to the fact that the non-interacting Hamiltonian anti-commutes with $C$. We use $\sigma_j$ and $\eta_j$ to denote Pauli matrices in band and valley space, respectively. 
}
\label{SymmetriesAndReps}\begin{ruledtabular}\begin{tabular}{cccc} 
$g$ & $c_{\vk}$ & $g^l_{\vk,\vk'}$ & $\xi_{\vk,\eta,\alpha}$ \\ \hline
$C_{2z}$ & $\eta_x c_{-\vk}$ & $\eta_x g^l_{-\vk,-\vk'} \eta_x$ & $\xi_{\vk,\eta,\alpha}= \xi_{-\vk,-\eta,\alpha}$ \\
$\Theta$ & $\eta_x c_{-\vk}$ &  $\eta_x(g^l_{-\vk,-\vk'})^* \eta_x$ & $\xi_{\vk,\eta,\alpha}= \xi_{-\vk,-\eta,\alpha}$ \\
$C_{2x}$ & $\sigma_zc_{(k_x,-k_y)}$ & $\sigma_zg^l_{C_{2x}\vk,C_{2x}\vk'}\sigma_z$ & $\xi_{\vk,\eta,\alpha}= \xi_{C_{2x}\vk,\eta,\alpha}$ \\
$C_{3z}$ & $c_{C_{3z}\vk}$ & $g^l_{C_{3z}\vk,C_{3z}\vk'} $ & $\xi_{C_{3z}\vk,\eta,\alpha}= \xi_{C_{3z}\vk,\eta,\alpha}$\\
$U(1)_v$ & $e^{i\varphi\eta_z} c_{\vk}$ &  $e^{-i\varphi\eta_z} g^l_{\vk,\vk'}e^{i\varphi\eta_z}$ & --- \\
$C$ & $\eta_z\sigma_yc_{\vk}$ & $\eta_z\sigma_yg^l_{\vk,\vk'}\eta_z\sigma_y$ & $\xi_{\vk,\eta,\alpha} = -\xi_{\vk,\eta,-\alpha}$
 \end{tabular}
\end{ruledtabular}
\end{center}
\end{table}

\begin{figure}[bt]
    \centering
    \includegraphics[width=\linewidth]{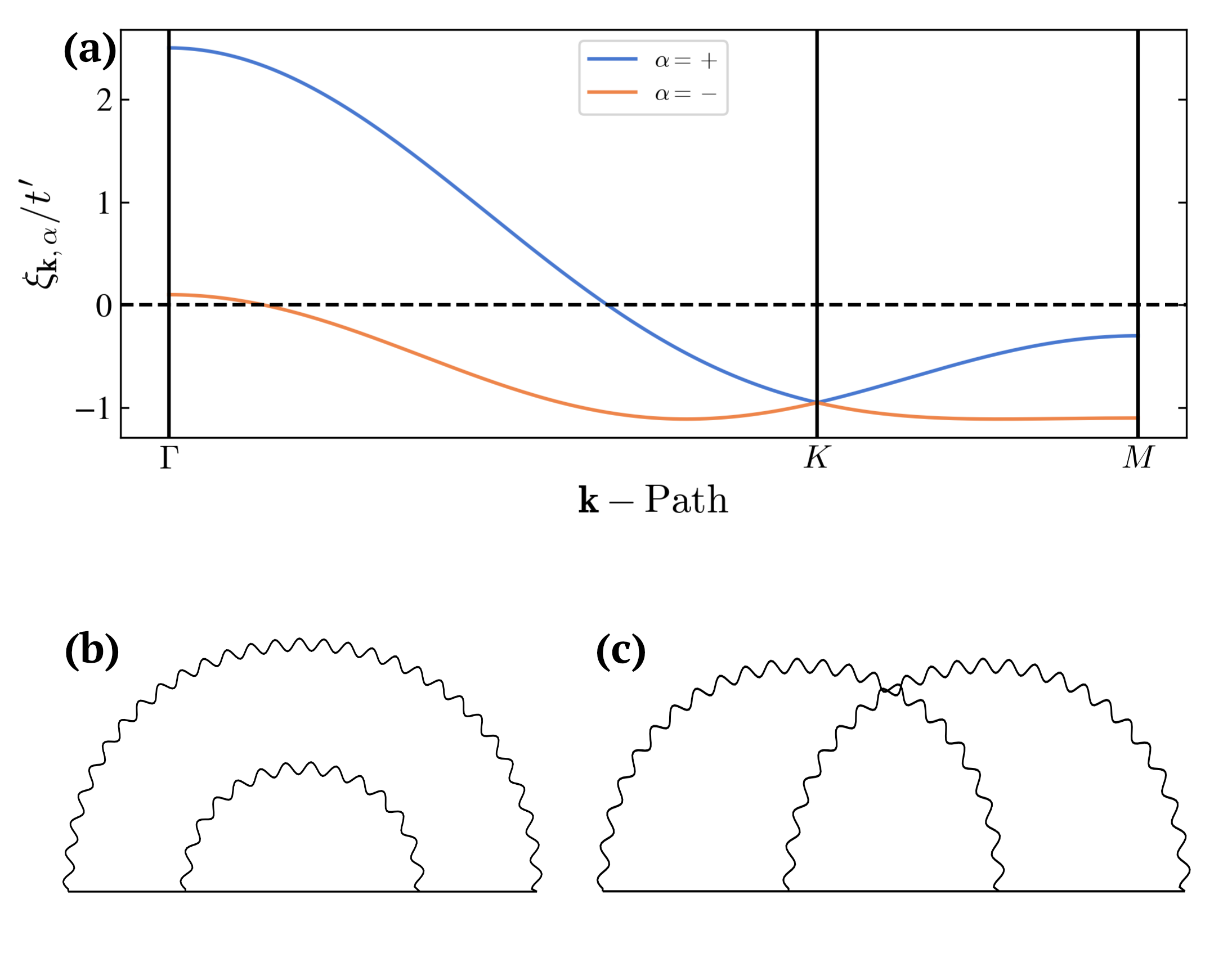}
    \caption{(a) Plot of the electronic band structure \eqref{eq:el_dispersion} along the $\Gamma-K-M$ path through the hexagonal Brillouin zone. The used parameters are $t/t'=0.8$ and $\mu/t'=0.4$. An example of a non-crossed (b) and a crossed (c) one-particle irreducible diagram.}
    \label{fig:Diagrams}
\end{figure}

In spite of the spin polarization in the normal state, there is still a spinless time-reversal symmetry $\Theta$ which we encode in the electronic dispersion by combining momentum and valley indices as $\xi_{\vk,\eta, \alpha}=\xi_{\eta\cdot\vk,\alpha}$; this also automatically ensures $C_{2z}$ symmetry. Additionally, we parameterize the dispersion via
\begin{equation}\label{eq:el_dispersion}
    \xi_{\vk,\alpha=\pm} = \pm \delta_\vk + \epsilon_\vk-\mu
\end{equation}
where the $\delta_{\vk}$ term represents the band-splitting and $\epsilon_\vk$ is introduced to capture the fact that the interaction-induced band renormalizations---related to the flavor-symmetry breaking in the normal state for $2<|\nu| <3$---break the chiral symmetry $C$ significantly. Both terms are constrained by $C_{3z}$ symmetry to obey $\delta_{C_{3z}\vk}=\delta_{\vk}$ and $\epsilon_{C_{3z}\vk}=\epsilon_{\vk}$. Since the $C_{2z}\Theta$ symmetry protecting the Dirac cones of single-layer graphene is not broken, neither by the moiré lattice nor by spin polarization, we still have Dirac cones now located at the K and K' points of the MBZ, leading to $\delta_{\text{K}} = \delta_{\text{K}'}=0$. To capture all of these properties in a minimal setting, we take $\delta_{\vect{k}}=t\vert1 + e^{\vect{a}_1\cdot \vk} + e^{-\vect{a}_2\cdot \vk}\vert$ and $\epsilon_{\vect{k}}=t'\sum_j\cos{(\vect{a}_j\cdot \vk)}$ with $t, t'\in \mathbb{R}$ and $\vect{a}_j = C^{j-1}_{3z} \left(\sqrt{3},0\right)^T$, $j=1,2,3$. The resulting dispersion along a one-dimensional cut through the MBZ is shown in \figref{fig:Diagrams}(a). As discussed in \refcite{BandOffDiagonalPairing}, a value of $t' \simeq t$, with $\sign (t') = -\sign (\nu)$ captures the qualitative features of the Hartree-Fock bands well. We note that the non-trivial topological character of the band structure is reflected by the different eigenvalues in the bands $\alpha = \pm $ under the valley preserving $C_{2x}$ symmetry \cite{PhysRevB.98.085435}, see \tableref{SymmetriesAndReps}.

Since the bosons will describe either optical phonons or gapped fluctuations of an uncondensed order parameter, we will choose fully gapped dispersions, $\omega^l_\vq> 0$. More specifically, to respect rotational symmetry, we take
\begin{equation}
    \omega^l_{\vect{q}}= \omega' + \omega_0\sum_j\cos{(\vect{a}_j\cdot \vq)},
\end{equation}
but we do not expect any of the main features of our results to crucially depend on the form of $\omega^l_\vq$. Throughout, we take $\omega_0/t' = 4$ and $\omega'/t'=6$.

In order to describe superconductivity within Eliashberg theory, we define the full normal and anomalous Green's functions as
\begin{subequations}\begin{align}
      \label{eq:def_G}
    G_{\alpha, \alpha'}(k,\eta)&=-\int_0^{\beta}d\tau e^{i \omega_n \tau}\langle T_\tau c_{\vk,\eta, \alpha}(\tau)c^{\dagger}_{\vk,\eta, \alpha'}(0)\rangle\\
    \label{eq:def_F}
    F_{\alpha, \alpha'}(k,\eta) &=-\int_0^{\beta}d\tau e^{i \omega_n \tau}\langle T_\tau c_{\vk,\eta, \alpha}(\tau)c_{{-\vk},-\eta, \alpha'}(0)\rangle
\end{align}\label{TheTwoPropagators}\end{subequations}
where $\beta=1/T$ denotes inverse temperature and the integral over imaginary time $\tau$ leads to the Matsubara frequency domain $i\omega_n=i (2n+1)\pi T$, $n\in\mathbb{Z}$. Further, $k=(i\omega_n, \vk)$ is a $2+1$-component vector containing Matsubara frequency in the first and momentum in the remaining two spatial components. 
The anomalous propagator in \equref{eq:def_F} has already been restricted to Cooper pairing with zero center-of-mass momentum, that is, between electrons of opposite valleys and momentum.
This form of pairing is generally favored by the time-reversal symmetry $\Theta$ as electrons with $(\vk, \eta)$ have the same energy as $(-\vk,-\eta)$. Furthermore, intravalley pairing and/or pairing between $\vk$ and $\vk'\neq -\vk$ are expected to be more fragile against impurities. 
In contrast, we do not assume anything about the coupling in band space, thus the propagators show a $2\times2$ structure in band space as a result of coupling inside as well as between the two electronic bands. 

The propagator of the bosonic field is given by
\begin{equation}\label{eq:def_D}
    D_l(q)=-\int_0^{\beta}d\tau e^{i \Omega_m \tau}\langle T_\tau a^{\pdag}_{\vq,l}(\tau)a^{\dagger}_{\vq,l}(0)\rangle
\end{equation}
where $a^{\pdag}_{\vect{q},l} = b^\dagger_{-\vect{q},l} + b^{\pdag}_{\vect{q},l}$. The bosonic Matsubara frequencies $i\Omega_m =i 2m\pi T$ with $m\in\mathbb{Z}$ are hidden in the first component of the $2+1$-component compound vector $q = (i\Omega_m, \vect{q})$.

The full electronic propagators in \equref{TheTwoPropagators} are related to their bare counterparts by self-energy and vertex corrections. In \figref{fig:Diagrams}(b,c), we show two such corrections which are fourth order in the fermion-boson coupling (\ref{ElPhonCoupl}). 
We follow standard Eliashberg techniques \cite{eliashberg1960, migdals_theorem} and neglect vertex corrections. This corresponds to neglecting the crossed diagram in \figref{fig:Diagrams}(c) while keeping the one shown in part (b). The remaining \textit{rainbow} diagrams can be summed up and captured conveniently in the normal and anomalous self energies:
\begin{align}
   \Sigma_{\alpha, \alpha'}(k,\eta)=T\sum_{k,\eta',\beta, \beta',l}&
    \left(g_{\vk, \vk'}^l\right)_{\alpha, \beta}^{\eta,\eta'}G_{\beta, \beta'}(k',\eta') \label{SigmaEquation}\\ \nonumber
    &\times\left(g_{\vk',\vk}^l\right)_{\beta', \alpha'}^{\eta',\eta}D_l(k-k')
\end{align}
\begin{align}
    \label{eq:def_phi}
    \Phi_{\alpha, \alpha'}(k,\eta)=T\sum_{k,\eta',\beta, \beta',l}&
    \left(g_{\vk, \vk'}^l\right)_{\alpha, \beta}^{\eta,\eta'}F_{\beta, \beta'}(k',\eta')\\ \nonumber
    &\times\left(g_{-\vk,-\vk'}^l\right)_{\alpha', \beta'}^{-\eta,-\eta'}D_l(k-k') 
\end{align}
In simpler one-band models, a common \cite{marsiglio_eliashberg_2020} simplification is to focus on the vicinity of the Fermi surface and to introduce a momentum-space-average at this point. However, since our main focus here is on the study of multi-band effects, such that it is \textit{a priori} unclear how the order parameter will behave in momentum space, we will keep the degrees of freedom in the entire MBZ and allow for full momentum dependence alongside with all other indices. We further assume that the central interaction-induced band renormalizations are already accounted for in $\xi_{\eta \cdot\vk,\alpha}$ and, thus, set $\Sigma=0$ in \equref{SigmaEquation}; instead, we focus on the superconducting instability described by the anomalous self energy $\Phi$.
Following the steps explicitly shown in \appref{app:nambu}, we find the anomalous propagator
\begin{align}\label{eq:expl_F}
    F_{\alpha, \alpha'}&(k,\eta=+) = \frac{1}{\theta(k)}\bigg[
    (i\omega_n-\xi_{\vk,-\alpha})(i\omega_n+\xi_{\vk,-\alpha'})\Phi_{\alpha, \alpha'}\\
    \nonumber
    &+\alpha\alpha' \Phi^*_{-\alpha, -\alpha'}(\Phi_{+,+}\Phi_{-,-}-\Phi_{+,-}\Phi_{-,+})
    \bigg]
\end{align}
with $\theta(k)$ denoting the determinant of the inverse Green's function in Nambu space given by \equref{eq:det_gf}. For brevity, we further suppressed the argument of $\Phi$, i.e., $\Phi_{\alpha,\alpha'} \rightarrow \Phi_{\alpha,\alpha'}(k,+)$ in \equref{eq:expl_F}. Together with \equref{eq:def_phi}, this leads to a self-consistency equation for $\Phi$, which can be regarded as the frequency-dependent generalization of the mean-field superconductor order parameter. This is the generalized Eliashberg equation that we solve to study multi-band superconductivity. It always has trivial solutions \ie $\Phi=0$ corresponding to the normal phase, and the superconducting regime is characterized by $\Phi \neq 0$.

\subsection{Intervalley phonons and T-IVC fluctuations}
So far, we have not specified the concrete physical meaning of the bosonic modes $b_{\vq,l}$ and what their coupling $g_{\vk, \vk'}^l$ is. Motivated by experiments  that reveal a particularly strong coupling of the flat-band electrons to intervalley phonon modes \cite{2023arXiv230314903C,2023arXiv230315551L}, our main focus will be on these phonon modes, which transform under $A_1$ and $B_1$. This is also in line with previous theoretical works on single-layer graphene \cite{intervalley_phonons_rg} showing that intervalley phonons are enhanced compared to the intravalley optical phonons. Focusing on the leading, $\vk$-independent, contribution to $g_{\vk, \vk'}^l$ and assuming that the Bloch states obey chiral symmetry $C$, one can show \cite{BandOffDiagonalPairing} by imposing the symmetries in \tableref{SymmetriesAndReps} (and noting that their microscopic coupling matrices are odd under $C$) that there is a unique coupling for each mode,
\begin{equation}
    g_{{\vk, \vk'}}^{A_1}=\lambda_{A_1}\eta_x\sigma_0 \quad \mathrm{and}\quad g_{{\vk, \vk'}}^{B_1}=\lambda_{B_1}\eta_y\sigma_0. \label{CouplingMatrices}
\end{equation}
Here, $\eta_{0,x,y,z}$ ($\sigma_{0,x,y,z}$) denote the Pauli matrices in valley (band) space and $\lambda_{A_1}, \lambda_{B_1}\in \mathbb{R}$ parametrize the coupling strengths. 

Since it is currently not known whether the pairing mechanism is driven by phonons or fluctuations of an electronic order parameter, we will also consider the latter possibility. Local probes \cite{TIVCExp} have revealed that the order parameter for the nearby correlated insulator is a $\Theta$-preserving intervalley coherent state (also known as T-IVC). Since the condensation of this order parameter breaks the $U(1)_v$ symmetry, it has two components. It turns out \cite{BandOffDiagonalPairing} that the coupling of these two components is again to leading order given by \equref{CouplingMatrices} and, hence, we can discuss intervalley phonons and T-IVC fluctuations simultaneously. 

For both of these pairing mechanisms, the two components, $l =1,2$, have to have an equal dispersion relation, $\omega_{\vq}^{l=1} = \omega_{\vq}^{l=2} \equiv \omega_{\vq}$ (and coupling constant), which follows from $U(1)_v$ symmetry. This simplifies the superconducting gap equation (\ref{eq:def_phi}) to 
\begin{equation}\label{eq:phi_sc_intervalley}
        \Phi_{\alpha, \alpha'}(k, \eta)=-T\sum_{k'}\lambda^2 F_{\alpha',\alpha}(k', \eta)D(k+k')
\end{equation}
where we have introduced the effective coupling constant $\lambda^2 \equiv \lambda_{A_1}^2 + \lambda_{B_1}^2$ and the phonon-mode-independent propagator $D(q) \equiv D_{A_1}(q) = D_{B_1}(q)$, which reads as
\begin{equation}
    D(q) = \frac{1}{i\Omega_m - \omega_{\vect{q}}} - \frac{1}{i\Omega_m + \omega_{\vect{q}}} <0.
\end{equation}

\subsection{Symmetries of gap equation}\label{sec:symmetries}
Applying the fermionic commutator relation to the anomalous propagator \eqref{eq:def_F} reveals the symmetry $F_{\alpha, \alpha'}(k,\eta) = -F_{\alpha',\alpha}(-k, -\eta)$
which is directly inherited by the order parameter as
\begin{equation}\label{eq:phi_sym_eta}
    \Phi_{\alpha, \alpha'}(k,\eta) = -\Phi_{\alpha',\alpha}(-k, -\eta).
\end{equation}
As this relations follows directly from the operator algebra it does not depend on any of the microscopic details. It proves especially useful in the numerics as it implies that $\Phi(\eta=+)$ and $\Phi(\eta=-)$ are not independent. Therefore we are able to only work with $\eta=+$ and drop the additional index from now on; we define $\phi_{\alpha, \alpha'}(k)\equiv \Phi_{\alpha, \alpha'}(k, \eta=+)$.

More symmetries can be revealed by analyzing the self-consistency equation itself. A common approach towards solving it is to linearize it in $\phi$, which is typically valid around the critical temperature: as long as the phase transition is continuous, the order parameter becomes arbitrarily small in the vicinity of this point and the linearization is well justified. Up to first order in $\phi$, 
\equref{eq:phi_sc_intervalley} can be written as
\begin{equation}\label{eq:phi_sc_lin_intervalley}
    \phi_{\alpha, \alpha'}(k) = \sum_{k',\beta, \beta'}
    M_{(k,\alpha,\alpha'),(k', \beta,\beta')} \phi_{\beta, \beta'}(k')
\end{equation}
where we absorbed all $\phi$-independent factors in the multi-index-space $(k, \alpha, \alpha')$ matrix $M$. Upon linearization, the self-consistency equation boils down to an eigenvalue equation, and we can analyze the symmetries by means of linear representation theory.
For the intervalley phonons the matrix elements are given by 
\begin{equation}\label{eq:gap_matrix_intervalley}
        M_{(k,\alpha, \alpha')(k',\beta, \beta')} = \frac{-T\lambda^2\delta_{\alpha,\beta'}\delta_{\alpha',\beta}D(k+k')}{(i\omega_{n'}-\xi_{{\vk'},\beta})(i\omega_{n'}+\xi_{{\vk'},\beta'})}.
\end{equation}
To discuss the symmetries of this matrix, we start by noting that $M$ is invariant under consecutive Matsubara frequency inversion and transposition in band-space \ie $M_{(\omega_n,\vk,\alpha,\alpha'),(\omega_{n'},{\vk'}, \beta,\beta')} = M_{(-\omega_n,\vk,\alpha',\alpha),(-\omega_{n'}, {\vk'}, \beta',\beta)}$. The operator associated with the joint Matsubara-inversion and band-transposition symmetry squares to unity such that the order parameter must fulfill $\phi_{\alpha, \alpha'}\left(i \omega_n,\vk\right)= \pm \phi_{\alpha', \alpha}\left(-i \omega_n,\vk\right)$. The physical origin of this symmetry is invariance of the system under $C_{2z}$, which acts as $\phi(i \omega_n,\vk)=\Phi(i\omega_n,\vk,+) \rightarrow \Phi(i\omega_n,-\vk,-) = -\Phi^T(-i\omega_n,\vk,+) = -\phi^T(-i \omega_n,\vk)$, where we used the fermionic anti-symmetry constraint in \equref{eq:phi_sym_eta}. In the numerics below, we will find that
\begin{equation}\label{eq:phi_sym_freq_band}
   \phi_{\alpha, \alpha'}\left(i \omega_n,\vk\right)= - \phi_{\alpha', \alpha}\left(-i \omega_n,\vk\right)
\end{equation}
is energetically favored, i.e., the order parameter is even under $C_{2z}$; this is intuitivedly expected, since the effective electron-electron interactions associated with the couplings in \equref{CouplingMatrices} will be an attractive intervalley coupling in the Cooper channel such that the superconductor order parameter is expected to have the same sign for opposite valleys (and opposite $\vk$).

It is common to have either even- or odd-frequency pairing, which we also recover in certain limits: if the band splitting is much larger than any superconducting energy scale, the order parameter will become entirely band diagonal such that \equref{eq:phi_sym_freq_band} implies odd-frequency pairing (while the corresponding $C_{2z}$-odd pairing would have even Matsubara parity). Furthermore, if the bands were degenrate (e.g., if $\alpha$ was spin and we had spin rotation invariance), $\xi_{\vk,+}=\xi_{\vk,-}$, $M$ in \equref{eq:gap_matrix_intervalley} would be invariant under $\omega_n \rightarrow -\omega_n$, leading to order parameters with definite Matsubara-frequency parity. It is only in the intermediate regime, where the bands are split yet sufficiently flat to allow for BOD components, where even- and odd-frequency pairing mixes, as determined by \equref{eq:phi_sym_freq_band}. 

Besides the above-mentioned $C_{2z}$ symmetry, $M$ additionally posses all of the remaining point symmetries of $D_6$. The solutions of the linear equation \eqref{eq:phi_sc_lin_intervalley} are eigenvectors of $M$ and thus either inherit the symmetries or occur in degenerate groups with the respective symmetry operation translating between the degenerate eigenvectors. More formally, the solutions will necessarily transform under one of the irreducible representations (IRs) of $D_6$; as a result of \equref{eq:phi_sym_freq_band}, this will be one of the $C_{2z}$-even IRs, $A_1$, $A_2$, or $E_2$.

Furthermore, as each Matsubara frequency in \equref{eq:gap_matrix_intervalley} is accompanied by a factor of $i$, inverting its sign is equivalent to complex conjugation. This anti-linear symmetry implies $\phi_{\alpha, \alpha'}\left(i \omega_n,\vk\right)= -e^{i\varphi} \phi^*_{\alpha, \alpha'}\left(-i \omega_n,\vk\right)$ with an arbitrary phase that can be gauged away by substituting $\phi\rightarrow e^{-i\varphi/2}\phi $. The extra minus sign was chosen such that a combination with \equref{eq:phi_sym_freq_band} leads to
\begin{equation}\label{eq:phi_sym_herm}
    \phi_{\alpha, \alpha'}\left(i \omega_n,\vk\right)= \phi^*_{\alpha', \alpha}\left(i \omega_n,\vk\right)
\end{equation}
which can be identified as time-reversal symmetry on the level of the order parameter.

Finally, there is another symmetry,
\begin{equation}\label{eq:phi_sym_aapbbp}
    M_{(k,\alpha, \alpha')(k',\beta, \beta')} = \alpha \alpha' \beta\beta' M_{(k,\alpha, \alpha')(k',\beta, \beta')},
\end{equation}
which implies 
\begin{equation}
    \sigma_z \phi(i\omega_n,\vk) \sigma_z = \pm \phi(i\omega_n,\vk). \label{SigmaZConstraint}
\end{equation}
Expanding $\phi$ in the Pauli-basis \ie $\phi(i\omega_n,\vk)=\sum_{\mu}f_\mu(i\omega_n, \vect{k})\sigma_\mu$, where $\mu=0,x,y,z$, we immediately see that there cannot be any mixing of diagonal and off-diagonal components. A positive (negative) sign in \equref{eq:phi_sym_aapbbp} implies a strictly (off)-diagonal order parameter. 
In the numerics we find a negative sign and thus off-diagonal solutions and only $f_{x,y} \neq 0$ while $f_{0,z}=0$.
\subsection{Numerical procedure}
We solve the linearized equation \eqref{eq:phi_sc_lin_intervalley} by analyzing the spectrum of the matrix $M$ given by \equref{eq:gap_matrix_intervalley}. This is done via explicitly constructing the matrix using a grid of $N_\vk=7\times 7$ points in the Brillouin zone and $N_m=38$ Matsubara frequencies chosen symmetrically around zero. For small temperatures we ensure a sufficient range of Matsubara frequencies by interpolating linearly spaced frequencies up to a certain cutoff $\Lambda$. This strategy was inspired by \refcite{Lewandowski2021}. The critical temperature $T_c$ is given by the highest temperature for which the spectrum of $M$ contains unity. We find it via a bisection method. At the critical temperature the \textit{leading} eigenvector solves the linear equation \eqref{eq:phi_sc_lin_intervalley}. 

This provides a good starting point for solving the full non-linear gap equation in \equref{eq:phi_sc_intervalley}. This is done via iteratively substituting the linear solution into the full equation. Explicitly this means
\begin{equation}
    \phi^{(i+1)} = \mathrm{RHS}\left(\phi^{(i)}\right)
\end{equation}
with $\mathrm{RHS(\cdot)}$ denoting the right hand side of \equref{eq:phi_sc_intervalley}. This approach works well in most temperature regimes and the trial-function converges towards a solution within $\sim 20$ iterations. However near the continuous phase transition the order parameter becomes small and the convergence process very slow. For this reason we modify the algorithm such that the candidate functions gets rescaled with an optimized factor $x\in [0.5,1.5]$ every $\sim 30$ iterations. With this modification the solutions converge well and we solve the equation on a grid of $51$ bandsplittings ($t$) $\times$ $625$ temperatures ($T$). We adjust the chemical potential $\mu$ such that the filling fraction stays constant for every value of $t$. In our convention $\tilde{\nu}=1$ corresponds to both bands being filled. Furthermore we fix the values $t'=1/2$ and $\lambda / t'=2\sqrt{15}$ throughout the entire calculation.

\section{Results for BOD pairing}\label{numericalresults}

\subsection{Solutions on imaginary axis}\label{sec:solutions_im_ax}
In the studied parameter regime, we two found candidate solutions from the leading eigenstates of the linearized multiband Eliashberg equation in \equref{eq:phi_sc_lin_intervalley}---a band-diagonal and a BOD solution, corresponding to the two different signs in \equref{SigmaZConstraint}. When iterating them in the non-linear equation in \equref{eq:phi_sc_intervalley}, only the latter BOD solutions remain while band-diagonal candidates converge towards the trivial solution $\phi=0$. As already anticipated above, the non-trivial solutions we obtain are even under $C_{2z}$, thus obeying \equref{eq:phi_sym_freq_band}. Taken together, we therefore have
\begin{equation}\label{eq:phi_c2z_even}
    \phi(i\omega_n, \vk) = f^{e}_{\vect{k}}(i\omega_n) \sigma_y + f^{o}_{\vect{k}}(i\omega_n) \sigma_x
\end{equation}
where $f^{e(o)}_{\vect{k}}$ is a real, even (odd) function in Matsubara space. This result closely connects to the $\sigma_y$ states found in \refcite{BandOffDiagonalPairing} on the mean-field level. We here go beyond mean-field and allow for Matsubara frequency dependence in the order parameter, which in turn leads to the admixture of an $\omega_n$-odd $\sigma_x$ contribution in \eqref{eq:phi_c2z_even} to the $\sigma_y$ term. Importantly, though, the symmetry in \equref{eq:phi_sym_aapbbp} still guarantees that no band-diagonal contribution can be admixed.

\begin{figure}[tb]
    \centering
    \includegraphics[width=\linewidth]{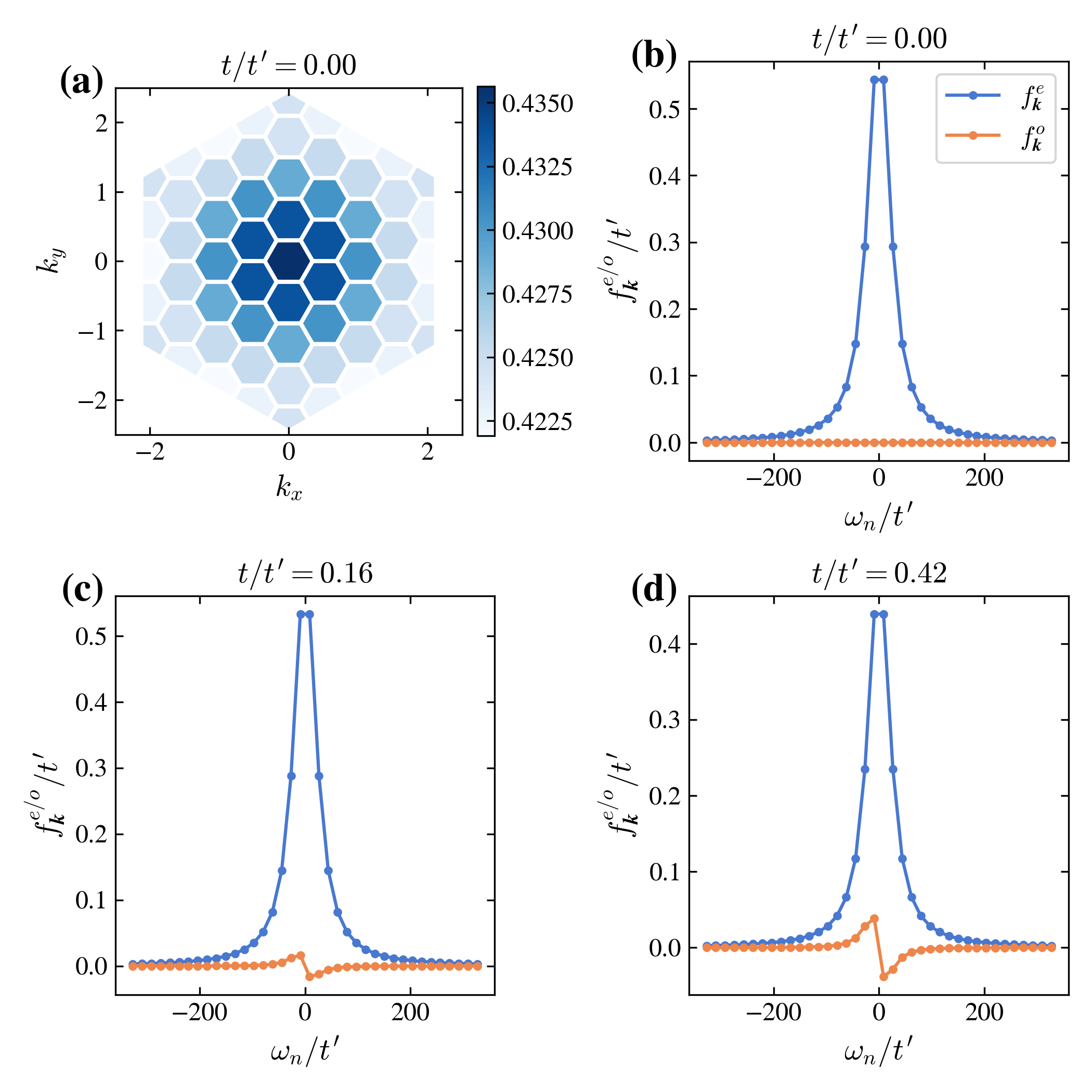}
    \caption{Solutions of the multiband Eliashberg equation \eqref{eq:phi_sc_intervalley} for the intervalley phonons. (a) shows the $(+,-)$ component of the order parameter at fixed Matsubara frequency $i\omega_n = i\pi T$ as a function of $\vk$ in the MBZ. The panels (b-d) show the frequency dependence of the even and odd components of $\phi$, defined in \equref{eq:phi_c2z_even}, at fixed momentum $\vk=\Gamma$ for different values of the band splitting $t/t'$.}
    \label{fig:Result1}
\end{figure}

To discuss the solutions more quantitatively, we start by analyzing the band-degenerate case $t=0$. Figure~\ref{fig:Result1}(a) shows the momentum dependence of the order parameter at fixed Matsubara frequency. It has no sign changes and transforms trivially under $C_{3z}$. In \cref{fig:Result1}(b), we plot the Matsubara-frequency dependence of the two terms in \equref{eq:phi_c2z_even} at fixed mometa; we observe that only one of them is finite, the $\omega_n$-even, $\sigma_y$ term. This was expected since, as explained above, $M$ in \equref{eq:gap_matrix_intervalley} becomes even in Matsubara frequency in the band degenerate case such that the solutions have to have definite Matsubara parity. Then the even-$\omega_n$ $\sigma_y$ state is naturally favored over the odd-$\omega_n$ alternative in the presence of an attractive interaction. 

As $\sigma_y$ anti-commutes with the representation of $C_{2x}$ in \tableref{SymmetriesAndReps}, we conclude from \cref{fig:Result1}(a) that the order parameter is odd under $C_{2x}$ and, thus, transforms under the IR $A_2$. As already pointed out in \refcite{BandOffDiagonalPairing}, the remarkable conclusion that electron-phonon coupling can induce a superconducting state transforming under a non-trivial IR \cite{PhysRevB.90.184512,PhysRevB.93.174509} is a result of the combination of broken spin-full time-reversal symmetry, the resultant inter-band pairing, and the topology-related representation of $C_{2x}$ in band space.

\begin{figure}[tb]
    \centering
    \includegraphics[width=1\linewidth]{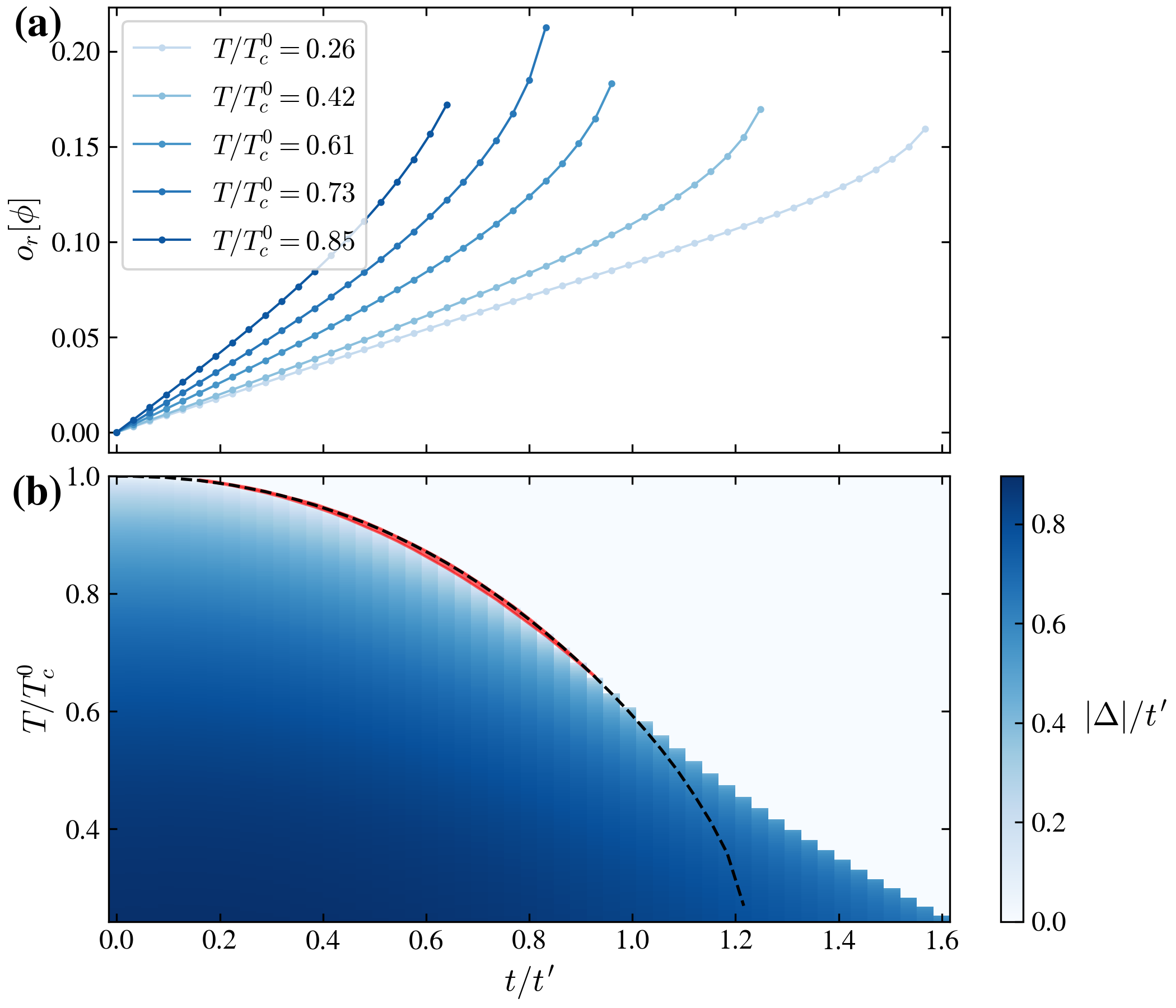}
    \caption{We plot the relative oddness (a), defined in \equref{eq:oddnes}, of the converged solutions as a function of bandsplitting for different temperatures. $T_c^0$ denotes the critical temperature of the zero splitting $t=0$ case. The different curves end at different values of $t/t'$ as a bigger band splitting generally suppresses the superconducting phase. The averaged absolute value of the order parameter as a function of bandsplitting and temperature is shown in panel (b). The black dashed line is the critical temperature found in the linearized gap equation. As the phase transition becomes first order above a critical value of $t'$,  it does not align with the actual critical temperature in that regime. The nodal region is indicated in red.}
    \label{fig:oddness_abs}
\end{figure}

The lower two panels of \cref{fig:Result1} show the frequency dependence of the solutions for non-zero band splitting, $t\neq 0$.  As the Matsubara-parity constraint is removed, the finite $\omega_n$-odd contribution in \equref{eq:phi_c2z_even} is admixed, $f^{o}_{\vect{k}} \neq 0$. To quantify the strength of the admixture, we define the relative oddness as
\begin{equation}\label{eq:oddnes}
    o_r[\phi] = \frac{\langle\vert f^{o}_{\vk}(i\omega_n)\vert\rangle_{\omega_n,\vk}}{
    \langle\vert f^{e}_{\vk}(i\omega_n)\vert+\vert f^{o}_{\vk}(i\omega_n)\vert\rangle_{\omega_n,\vk}
    }
\end{equation}
and plot it as a function of bandsplitting and temperature in \cref{fig:oddness_abs}(a). This representation shows that the odd contributions increase monotonously as function of bandsplitting. Additionally we note that away from $t=0$ the relative oddness increases as the temperature gets lowered. In the band-degenerate case the solutions have a well-defined even parity for all temperatures ($o_r=0$).

To obtain a measure, $\Delta$ for the strength of the order parameter, we evaluate $\phi$ at the lowest Matsubara frequency and average the absolute value over all other indices. This specific way of reducing the multicomponent order parameter to a single number will prove useful in the analysis of the excitation spectrum in the next section. We plot the absolute value, $|\Delta|$, in \cref{fig:oddness_abs}(b). For small bandsplittings the order parameter clearly goes to zero in a continuous way. However, it is rather remarkable that this transition becomes discontinuous towards the right-hand side of the plot ($t/t' \gtrsim 0.81$) as the band splitting gets bigger. This means that the order parameter is never infinitesimally small in this regime and one needs to be especially careful when working with the linearized equation. To emphasize this, the dashed black line in \cref{fig:oddness_abs}(b) represents the critical temperature obtained by solving the linearized gap equation. We can see that, in the strong-splitting regime, non-trivial solutions of the full non-linear gap equation exist way beyond the dashed line.

\subsection{Excitation spectrum}\label{sec:ex_spectrum}
A prominent feature of the superconducting phase is the gapped excitation spectrum of the quasiparticles. On the mean-field level, a sharp spectrum is obtained by diagonalizing the Bogoliubov de Gennes (BdG) Hamiltonian. This requires a static order parameter which can be recovered from the frequency dependent dynamic order parameter $\phi$ by evaluating it at $\omega=0$. As we are dealing with fermionic Matsubara frequencies we approximate this by using the lowest positive frequency and find the \textit{almost} static mean-field order parameter
\begin{equation}
    \Delta(\vk)= \phi\left(i\omega_0=i\pi T, \vk\right) \label{eq:meanfieldDeltaDef}
\end{equation}
which then shares the off-diagonal structure in band space with $\phi$. This leads to rather unusual physical effects and an extensive analysis on the mean field level can be found in \refcite{BandOffDiagonalPairing}.
The resulting BdG spectrum is given by the four bands $\pm E_{\vk, \pm}$ with
\begin{equation}\label{eq:bdg_sig_y}
    E_{\vk,\pm}=\pm\vert\delta_\vk\vert + \sqrt{(\epsilon_\vk-\mu)^2 + \vert\Delta_\vk\vert^2}.
\end{equation} 
The expression differs from the usual $\sqrt{\xi^2 + \vert\Delta\vert^2}$ spectrum found in a simpler one-band model. A prominent feature is the relative minus sign between the two positive terms in \equref{eq:bdg_sig_y}. This means that $\vert\Delta_\vk\vert\neq0$ does not necessarily imply a gapped excitation spectrum. This nodal region is indicated in red in \cref{fig:oddness_abs}(b). While this region is small and the superconductor's BdG spectrum is technically fully gapped in a large part of the phase diagram, the off-diagonal nature still has important consequences for the spectral function in the gapped BdG regime since the excitation gap is much smaller than the order parameter. 

\begin{figure}[bt]
    \centering
    \includegraphics[width=1\linewidth]{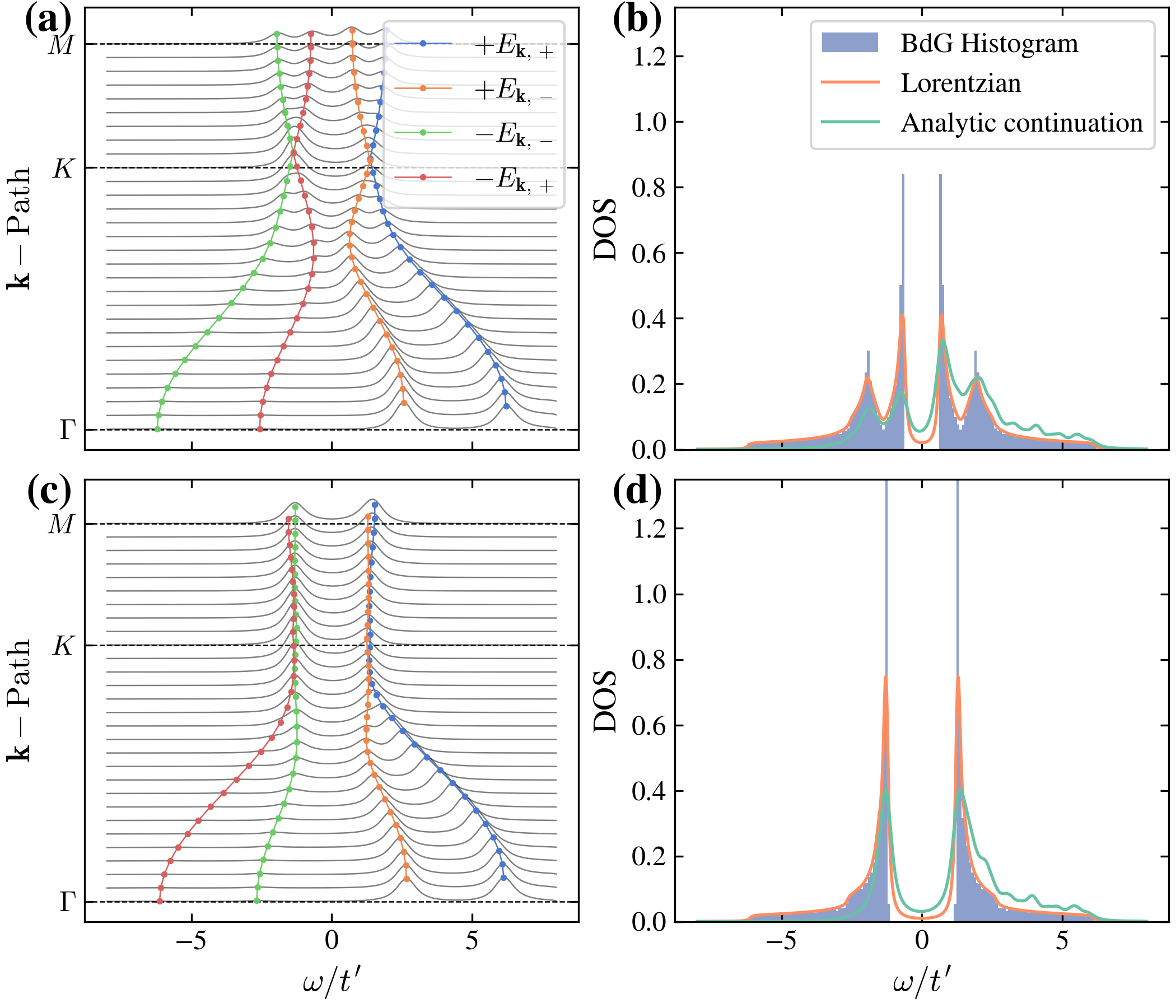}
    \caption{Waterfall plots consisting of the analytically continued spectral functions with offset corresponding to different momenta along the $\Gamma-K-M$ path for the BOD (a) and band-diagonal (c) order parameter. The blue, orange, green and red lines are the four bands of the BdG spectrum. The density of states is calculated using three different methods and shown in (b) for the BOD and in (d) for the band-diagonal order parameter. The blue bars show a histogram (with $200$ bins) of the BdG spectrum. The orange curve accounts for the finite lifetime by approximating the DOS as a momentum sum over Lorentzians centered around the BdG spectrum. The green curve is calculated by summing the analytically continued spectral functions over all momenta. The DOS plots are normalized such that the total area under the curves is equal to $1$. The temperature and band splitting is $T=0.95 T_c$ and $t/t'=0.61$, respectively, in all plots.}
    \label{fig:ac_sigy_0.95}
\end{figure}

To see this, we next compute the spectral function taking into account the full frequency dependence of $\phi$. We construct the interacting electronic Green's function \eqref{eq:def_G} and numerically performing an analytic continuation back to the real axis. We use a maximum entropy method implemented by \refcite{kaufmann2021anacont}. The resulting spectral functions \ie the imaginary part of the retarded Green's functions are represented as a waterfall plot along the $K-\Gamma-M$ path in figure \cref{fig:ac_sigy_0.95}(a). The four colored curves are the BdG bands. The bands clearly trace out the peaks of the spectral functions, showing that the mean-field picture based on \equref{eq:meanfieldDeltaDef} and the numerical analytical continuation yield compatible results. The height and width of the peaks carry additional information about the quasiparticle weight and lifetime. Note, however, as the analytic continuation is performed on the basis of finitely many Matsubara frequencies, the interpretation of the exact shapes of the curves is limited.

As a consistency check and to obtain a state to compare our BOD superconductor with, we also considered the pairing states resulting from a bosonic mode with coupling function $g=\lambda \sigma_x \eta_0$, $\lambda \in \mathbb{R}$, in \equref{ElPhonCoupl}. Solving the self-consistency equation results in a band-diagonal [$+$ sign in \equref{SigmaZConstraint}] order parameter, which is even in $\omega_n$ [and, thus, $C_{2z}$-odd, see discussion above \equref{eq:phi_sym_freq_band}]. There are further no sign changes as a function of momentum $\vk$, such that the pairing state transforms under the IR $B_1$. This behavior was expected because the boson mediates an attractive intra-valley interaction, which favors a constant, band-diagonal order parameter that changes sign \textit{between} the two valleys [due to the Fermi Dirac constraint in \equref{eq:phi_sym_eta}].

We apply the same numerical procedures as above to the $B_1$ state and show the resultant waterfall plot and the DOS in \cref{fig:ac_sigy_0.95}(c) and (d). As all parameters as well as the numerical methods are the same for both interactions, we can directly compare the plots. This allows us to eliminate possible effects of the uncontrolled analytic continuation and isolate the effects originating in the BOD nature of the order parameter. We see that the band-diagonal order parameter clearly opens an excitation gap while the BOD order parameter allows e.g.~the two $\pm E_{\vect{k},-}$ bands to get close between the $\Gamma$ and $K$ point. This results in an excitation gap that is smaller than $\vert\Delta_\vk \vert$ even before entering the nodal region. 

Finally, to generalize our analysis, we also looked into different forms of the coupling vertices $g_{\vk,\vk'}^l$ in \equref{ElPhonCoupl} of the bosonic modes.  We observed particularly exotic behavior when allowing both inter- and intra-band scattering for intravalley modes, $g_{\vk,\vk'}^l \propto (\sigma_0 + \sigma_x) \eta_0$. Solving the Eliashberg equations in such a system reveals an order parameter that combines band-diagonal and BOD components. As a result, each particle-like band tends to avoid crossings with all hole-like bands, and vice versa. This creates multiple, opposing repulsive ``forces'' between the bands that, depending on the parameters, lead to intriguing behavior when analyzing the excitation gap as a function of the magnitude of the order parameter. A detailed analysis is provided in \appref{sec:broken_chiral_sym}.

\section{Superfluid Stiffness}\label{SuperfluidStiffness}
In this section, we analyze the superfluid stiffness of a superconductor with BOD order parameter, accounting for quantum geometric effects. To this end, we follow the derivation presented in \refcite{PhysRevB.95.024515} which is valid for multiband systems, within mean-field theory.
To apply this formalism and to transparently encode the geometry of the normal-state Bloch bands, we will work in a different basis than in the other sections; we choose a single-particle basis where the superconducting order parameter is diagonal in the internal degrees of freedom (again represented by Pauli matrices $\sigma_j$), which can be thought of as two sublattices giving rise to the two bands we study pairing in.
The mean-field BdG Hamiltonian in this basis read as
\begin{align}\begin{split}\label{eq:H_BdG_maintext}
    &H_\bdg = \\
    &\begin{pmatrix}
        \xi_\vk \sigma_0 + \delta^{(1)}_\vk \sigma_1 + \delta^{(2)}_\vk\sigma_2 & \Delta_\vk \sigma_0 \\
        \Delta_\vk^* \sigma_0 & -\xi_\vk \sigma_0 + \delta^{(1)}_\vk\sigma_1 + \delta^{(2)}_\vk\sigma_2  \\
    \end{pmatrix}.
\end{split}\end{align}
It is straightforward to verify that the superconducting order parameter is completely off-diagonal when transforming \equref{eq:H_BdG_maintext} to the band basis. Furthermore, when writing $\delta^{(1)}_\vk + i\delta^{(2)}_\vk = \delta_\vk e^{i\theta_\vk}$, with $\delta_\vk > 0$, one easily verifies that the spectrum of \equref{eq:H_BdG_maintext} is precisely \equref{eq:bdg_sig_y}. Additionally, we account for the potential momentum dependence of the order parameter giving rise to terms involving momentum derivatives of $\Delta_\vk=|\Delta_\vk|e^{i\varphi_\vk}$, which will become relevant later in this section. We find the superfluid stiffness as
\begin{widetext}
\begin{align}\label{eq:stiffness}
    D_s^{i j}&=\sum_{\vk, \pm}
    \left[\left(\partial_i\xi_\vk\right)\left(\partial_j \xi_\vk\right)-\left(\partial_i \delta_\vk\right)\left(\partial_j \delta_\vk\right)+\left(\partial_i |\Delta_\vk|\right)\left( \partial_j|\Delta_\vk|\right)\right] 
    \frac{|\Delta_\vk|^2}{\xi_\vk^2+|\Delta_\vk|^2} \\ \nonumber
    \times& \Bigg[
    \frac{1}{{\sqrt{\xi_\vk^2+|\Delta_\vk|^2}}}\tanh{\left(\frac{\beta}{2} E_{\vk,\pm}\right)}
    -\frac{\beta}{2} \frac{1}{\cosh ^2\left(\frac{\beta}{2}E_{\vk, \pm}\right)}\Bigg]  \\\nonumber
    +&\left(\partial_i \theta_\vk\right)\left(\partial_j \theta_\vk\right) \frac{|\Delta_\vk|^2}{\xi_\vk^2+|\Delta_\vk|^2}\Bigg[
    \pm\delta_\vk{\tanh{\left(\frac{\beta}{2} E_{\vk,\pm}\right)}}
    -\delta_\vk^2\frac{\tanh \left(\frac{\beta}{2}E_{\vk, \pm}\right)}{E_{\vk, \pm}}\Bigg] \\
    -& \left[\left(\partial_i |\Delta_\vk|\right)\left( \partial_j|\Delta_\vk|\right) + |\Delta_\vk|^2\left(\partial_i \varphi_\vk\right)\left(\partial_j \varphi_\vk\right)\right]\frac{1}{\sqrt{\xi_\vk^2 + |\Delta_\vk|^2}}\tanh{\left(\frac{\beta}{2} E_{\vk,\pm}\right)}\nonumber,
\end{align}
\end{widetext}
where $\partial_i \equiv \partial/\partial k_i$. The first term includes derivatives of the dispersions and is independent of the phase angles $\theta_\vk$ and $\varphi_\vk$. The second term is proportional to derivatives of $\theta_\vk$, which accounts for how the eigenstates of the Hamiltonian change as a function of the orientation of $(\delta_\vk^{(1)}, \delta_\vk^{(2)})^T$ and, as such, encodes the quantum geometry of the normal state. 
In the band-degenerate limit, $\delta_\vk \rightarrow 0$ these geometric contributions naturally vanish in our model, and the BdG dispersion simplifies to the familiar form $E_{\vk,\pm} = \pm |\delta_\vk \vert +\sqrt{\xi_\vk^2 + |\Delta_\vk|^2} \rightarrow E_\vk=\sqrt{\xi_\vk^2 + |\Delta_\vk|^2}$. In this edgecase, the sum over $\pm$ in \equref{eq:stiffness} simply yields an extra factor of two, and the result resembles a single-band system. The last term involves derivatives of $\varphi_\vk$ and therefore describes how a potentially momentum-dependent superconducting order parameter induces additional variations of the BdG Bloch states across the Brillouin zone. 

\begin{figure}
    \centering
    \includegraphics[width=1\linewidth]{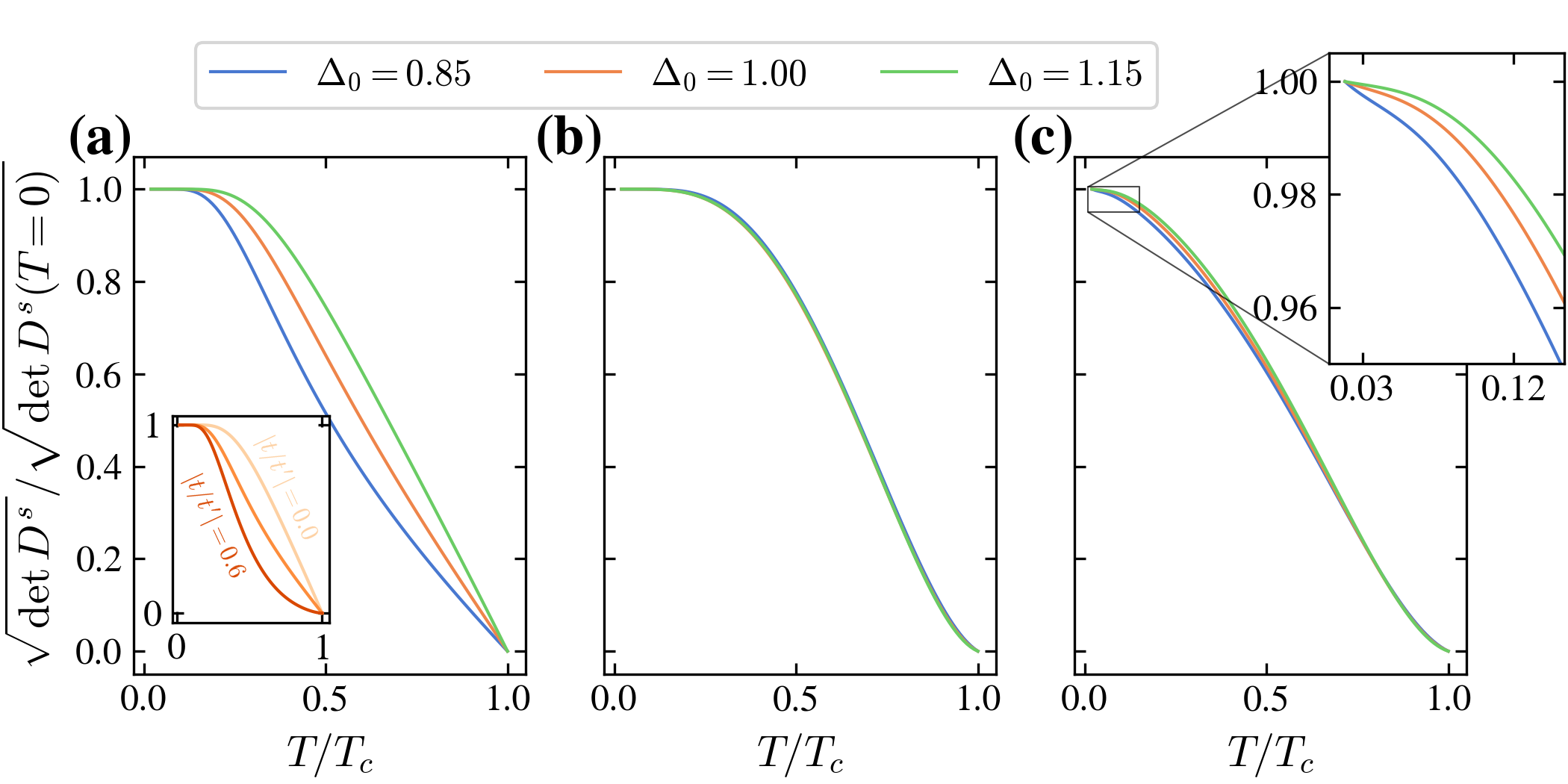}
    \caption{
    Panel (a) shows the superfluid stiffness as a function of temperature calculated for an $A_2$ BOD pairing state with $\vec{k}$-independent order parameter of different overall strength $\Delta_0$ ($\Delta_0=1$ corresponds to the BCS relation $\Delta(T=0) = 1.764T_c$). Here the chemical potential is set to $\mu=2.5t$. In the inset we fix $\Delta_0=1$ and vary $t/t'$.For the gapped / nodal nature of the BdG excitation spectrum the value of $\mu$ is of no relevance for the $\vec{k}$-independent order parameter. In panel (b) we plot the same quantity for a chiral order parameter in the fully gapped regime, at $\mu=-2.16t$. At $\mu=2.5t$ the system is in the nodal regime and we plot the superfluid stiffness in panel (c). The inset shows a low temperature zoom of the non-saturating behavior close to $T=0$. The remaining parameters are $T_c=2.5t$, $t'=-2.2t$, $t>0$ in all plots.}
    \label{fig:stiffness}
\end{figure}

We start the discussion by modeling the entries of the BOD $A_2$ order parameter as $\Delta_\vk=\Delta^{s} = 1.764 T_c \Delta_0\tanh \left(1.74 \sqrt{\frac{T_c}{T}-1}\right)$ with critical temperature $T_c$ and $\Delta_0\in\mathbb{R}$. The superscript emphasizes the momentum-independent $s$-wave nature of the order parameter. Here, all terms involving momentum derivatives of $\Delta_\vk$ vanish. Figure \ref{fig:stiffness}(a) shows the superfluid stiffness as a function of temperature for different values of $\Delta_0$. When $\Delta_0=1$ we recover the BCS relation $\Delta(T=0)/T_c = \pi e^{-\gamma}  \approx 1.764$, with Euler–Mascheroni constant $\gamma$.
First, we note that the scale of the order parameter, $\Delta_0$, influences the shape of the curves. For smaller order parameter magnitudes, we obtain an inflection point---not unlike experiment \cite{banerjee_superfluid_2024}. Second, since the superconductors for all curves in \figref{fig:stiffness}(a) are fully gapped at low temperatures, one would expect that $D_s(T)-D_s(T=0)$ is exponentially suppressed at low $T$. This is indeed visible, however, compared to conventional band-diagonal superconductivity ($t=0$ limit of our theory), the temperature at which the saturation sets is reduced with increased band splitting [see inset of \figref{fig:stiffness}(a)].

There is an additional term beyond \equref{eq:stiffness} arising from the multi-band nature of the system that needs to be discussed: as pointed out in \refcite{huhtinen_revisiting_2022}, while the (constant) order parameter in a single-band system can always be chosen real, even at nonzero Cooper pair momentum $\vq$, this does not hold for systems with multiple bands. By a global phase shift only one of the two components of the order parameter can be chosen real. In \appref{app:stiffness_correction}, we discuss the resulting additional terms [see \equref{eq:stiffness_corr}] as well as how the terms can be computed for an $s$-wave order parameter. A numerical evaluation shows that, in our case, the terms stated in \equref{eq:stiffness} are dominant and the remaining terms yield corrections at most of the order of $10^{-4}$ compared to the main contributions in our results. We will therefore neglect this term in the following. 

Besides $\vec{k}$-independent $A_2$ BOD pairing, we also discuss the stiffness for chiral BOD superconductivity. While a (predominantly) band-diagonal chiral order parameter is know to give rise to a fully developed gap (except for possible isolated points in parameter space), it was shown in \refcite{BandOffDiagonalPairing} that chiral BOD pairing yields extended nodal regimes; in fact, even for just a fixed order parameter, one finds gapped to nodal transitions as a function of filling, which could explain the tunneling data \cite{Oh_2021,TunnelingPerge}. This is our motivation to study the stiffness of such a state as well.

We model it as $\Delta^{\mathrm{c}}_\vk(T)= \Delta^{s}(T) \chi_\vk$. The momentum dependent part, $\chi_\vk = X_\vk + i \,Y_\vk$ is constructed as $X_\vk =X_\vk^o + \eta X_\vk^e$, $Y_\vk =-2/\sqrt{3}\left( X_{\vk} + X_{C_{3z}^{-1}\vk}\right)$ with odd and even basis functions \cite{BandOffDiagonalPairing}
\begin{align*}
    X_\vk^{{o}} &= -\frac{4}{3}\left[\sin \left(\vect{a}_1 \vk\right)-\sin \left(\vect{a}_2 \vk\right)\right]\\
    X_\vk^{{e}} &= -\frac{4}{3}\left[\cos \left(\vect{a}_1 \vk\right)+\cos \left(\vect{a}_2 \vk\right)-2 \cos \left(\left(\vect{a}_1+\vect{a}_2\right) \vk\right)\right],
\end{align*}
respectively. We set $\eta=0.2$ for concreteness. To ensure that $\Delta^\mathrm{s}$ and $\Delta^c$ are directly comparable, we additionally rescale $\chi_\vk$ such that the the Brillouin zone average is $\langle |\chi_\vk|\rangle_{\vk\in\mathrm{BZ}}=1$. As opposed to the $\vec{k}$-independent case above, the momentum derivatives of the order parameter are now nonzero, introducing additional quantum geometric effects, originating from the order parameter [last term in \equref{eq:stiffness}]. Although the spectrum of \equref{eq:H_BdG_maintext} is independent of $\varphi_\vk$, the eigenstates change with the orientation of $(X_\vk, Y_\vk)^T$, adding another layer of quantum geometric contributions.

Another important difference is that, unlike $A_2$ BOD pairing where the presence or absence of a gap depends on the magnitude of the order parameter, the chiral BOD state can have BdG Fermi surfaces at arbitrary strength of the order parameter.
The criterion for such surfaces to appear is simply the existence of a point $\vk_j \in \mathrm{BZ}$ such that $\Delta_{\vk_j}=0$ and $\vert \delta_{\vk_j}\vert>\vert\xi_{\vk_j}\vert$. As long as $\delta_\vk$ features Dirac cones, the opposite inequality will be fulfilled at those points, \ie $\vert\delta_\vk\vert \leq \vert\xi_\vk\vert < \sqrt{\xi_\vk^2 + |\Delta_\vk|^2}$ for any $\Delta_\vk \neq 0$. By continuity, this implies that there must exist a point $\vk^*\in\mathrm{BZ}$ where $\vert \delta_{\vk^*}\vert = \sqrt{\xi_{\vk^*}^2 + |\Delta_{\vk^*}|^2}$, \ie a sign change of $E_{\vk,-}$ and thus a BdG Fermi surface \cite{BandOffDiagonalPairing}. Whether this criterion is fulfilled is determined by the value of the chemical potential, yielding a mechanism to tune between the gapped and nodal phase as already mentioned above.

In the gapped phase, similar to conventional $s$-wave pairing, $T$ becomes the smallest energy scale at low temperatures, and all terms in \equref{eq:stiffness} saturate close to $T=0$. This scenario is shown in \figref{fig:stiffness}(b). Qualitatively it displays the same features as the $A_2$ BOD pairing state, however the terms are much less susceptible to varying $\Delta_0$.
In contrast, in the nodal regime the low-temperature asymptotics of $D_s^{ij}$ changes significantly. We show this in \figref{fig:stiffness}(c). Here, the BdG spectrum stays nodal all the way to $T=0$ and thus $E_{\vk, -}/T$ stays finite for some $\vec{k}$ close to $\vk^*$. Generally, this leads to non-trivial scaling behavior and prevents the curves from saturating close to $T=0$. The scaling analysis presented in \appref{sec:app_low_temp_scaling} shows universal, quadratic contributions for small temperatures; however, there is an additional term adding non-universal contributions, scaling as $\log(T)$, which only becomes dominant at exponentially low temperatures $T$. It generally leads to a dip in the superfluid stiffness for $T\rightarrow 0$, as the logarithmic contribution diverges and dominates the other terms. Numerically, we did not resolve those effects all the way down to the smallest considered temperature, $0.02T_c$. At such low energy scales, other phenomena, such as disorder, are likely to play a significant role, complicating the interpretation and raising questions about the physical and experimental relevance of the logarithmic term in this regime.
In the intermediate temperature range, we note that $D_s$ generally features less curvature relative to the gapped regime. Depending on the parameters, the nodal chiral state can feature a large linear regime which does not show any signs of exponential saturation. This makes it a potential candidate for recent experimental observations \cite{banerjee_superfluid_2024}.

\section{Conclusion}\label{Conclusion}
In this work, we investigate multiple aspects of band-off-diagonal superconductivity as a candidate pairing state for twisted multilayer graphene systems. We employ a minimal two-band model and induce intervalley pairing by electron-phonon coupling or fluctuations of a time-reversal symmetric intervalley-coherent order parameter. We incorporate interband effects by avoiding restrictions on the band-space structure of the order parameter and account for these effects beyond mean-field theory using an Eliashberg theory framework. This generalized framework also incorporates phonon dynamics, resulting in a Matsubara-frequency-dependent order parameter that takes the form of a matrix in band space. By deriving (\secref{sec:general_theroy}) and solving (\secref{sec:solutions_im_ax}) a self-consistency equation for the order parameter, we demonstrate that the system exhibits pure band-off-diagonal superconductivity, without any admixed band-diagonal components.
Furthermore, a symmetry analysis in \secref{sec:symmetries} reveals a coupling between band and Matsubara-frequency symmetries. This interplay implies that the order parameter does not exhibit a well-defined Matsubara parity in the general case.

We analyze the solutions over a range of temperatures and band splittings, observing that larger band splittings generally lead to stronger odd-frequency contributions. Within the explored parameter space, we find that interband Cooper pairs are favored, and increasing the band splitting tends to lower the critical temperature, eventually suppressing the superconducting phase entirely.
To study the excitation spectrum, we numerically compute the  spectral functions via analytic continuation to the real axis (\secref{sec:ex_spectrum}). A prominent feature of band-off-diagonal superconductivity is the significant reduction of the excitation gap relative to the absolute value of the order parameter. A comparison with mean-field spectra shows overall consistency with these findings. We also explore the behavior of more exotic multi-band pairing scenarios (\appref{sec:broken_chiral_sym}).

To further characterize the properties of the band-off-diagonal superconductivity studied in this work, we computed the superfluid stiffness as a function of temperature for both a $\vec{k}$ independent and a chiral superconducting state (\secref{SuperfluidStiffness}). Our expression incorporates corrections proposed by \refcite{huhtinen_revisiting_2022}, which, however, were found to be negligible in this context. 
Our results show that the reduction of the gap of a $\vec{k}$ independent band-off-diagonal state can lead to a significant suppression of the temperature scale below which the stiffness saturates exponentially. In contrast, the chiral band-off-diagonal state displays a persistent nodal regime, even at low temperatures. 
This leads to a distinct lack of saturation in the superfluid stiffness.

\begin{acknowledgments}
B.P. acknowledges discussions with C.~Bühler and S.~Banerjee. M.S.S. thanks M.~Christos and L.~Classen for discussions and comments. B.P.~and M.S.S.~acknowledge funding by the European Union (ERC-2021-STG, Project 101040651---SuperCorr). Views and opinions expressed are however those of the authors only and do not necessarily reflect those of the European Union or the European Research Council Executive Agency. Neither the European Union nor the granting authority can be held responsible for them.

\end{acknowledgments}

\bibliography{draft_Refs}

\onecolumngrid

\begin{appendix}

\section{Nambu Formalism}\label{app:nambu}
Expanding the propagators (\ref{eq:def_G},\ref{eq:def_F}) in the selfenergies can be done very elegantly by introducing the Nambu-Gor'kov Green's function
\begin{equation}
    \mathcal{G}(k,\eta) = 
    \begin{pmatrix}
        G(k,\eta) & F(k,\eta)\\
        F^\dagger(k,\eta) & -G(-k,-\eta)
    \end{pmatrix}
\end{equation}
together with the matrix-valued selfenergy
\begin{equation}
    \hat{\Sigma}(k,\eta) = 
    \begin{pmatrix}
        \Sigma(k,\eta) & -\Phi(k,\eta)\\
        -\Phi^\dagger(k,\eta) & -\Sigma(-k,-\eta)
    \end{pmatrix}.
\end{equation}
Withing this framework the matrix multiplication algebra accounts for all the mixing effects and the inverse interacting propagator can be written in the familiar way
\begin{equation}\label{eq:gor_dyson_expansion}
    \mathcal{G}^{-1}(k,\eta) = \mathcal{G}_0^{-1}(k,\eta)-\hat{\Sigma}(k,\eta)
\end{equation}
where 
\begin{equation}
    \mathcal{G}_0^{-1}(k,\eta)=\begin{pmatrix}
        i\omega_n - \xi_{\eta\vk, +} & 0 &0 &0 \\
        0&i\omega_n - \xi_{\eta\vk, -}&0&0\\
        0&0& i\omega_n + \xi_{(-\eta)({-\vk}), +}&0\\
        0&0&0& i\omega_n + \xi_{(-\eta)({-\vk}), -}
    \end{pmatrix}
\end{equation}
denotes the non-interacting Gor'kovs Green's function. Setting $\Sigma=0$, explicitly inverting equation \eqref{eq:gor_dyson_expansion} and extracting $F=\mathcal{G}_{1,2}$ as the upper right component of the Gor'kov Green's function leads to equation \eqref{eq:expl_F}. The determinant of the inverse Gor'kov Green's function is given by
\begin{align}\label{eq:det_gf}
    \theta(k,+)=& (\omega_n^2+\xi_{\vk,+}^2)(\omega_n^2+\xi_{\vk,-}^2) \\
    \nonumber
    +&(\omega_n^2+\xi_{\vk,-}^2)\vert\Phi_{+,+}\vert^2 + (\omega_n^2+\xi_{\vk,+}^2)\vert\Phi_{-,-}\vert^2\\
    \nonumber
    -& (i\omega_n - \xi_{\vk,+})(i\omega_n + \xi_{\vk,-})\vert\Phi_{-,+}\vert^2
    - (i\omega_n - \xi_{\vk,-})(i\omega_n + \xi_{\vk,+})\vert\Phi_{+,-}\vert^2\\
    \nonumber
    +& \vert\Phi_{+,+}\Phi_{-,-} - \Phi_{+,-}\Phi_{-,+}\vert^2
\end{align}
where all $\Phi$ are evaluated at $(k,\eta=+)$. Due to equation \eqref{eq:phi_sym_eta} it is sufficient to only regard the $\eta=+$ case.

\section{Other multi-band superconductors}\label{sec:broken_chiral_sym}

In this appendix, we examine some additional interesting behavior resulting from a scattering vertex of the form
\begin{equation}
    g_{\vect{k,k'}}^l = \lambda (\sigma_0 + \sigma_x) \eta_0
\end{equation}
with $\lambda\in\mathbb{R}$. Solving the resulting Eliashberg equation using the same method as described in the main text we find an order parameter of the form
\begin{equation}
    \phi\left(i\omega_n, \vk\right) = f\left(i\omega_n, \vk\right) 
    \left(\sigma_0 + \sigma_x\right)
\end{equation}
where $f\left(i\omega_n, \vk\right)$ transforms trivially under the combined lattice and Matsubara symmetry group, \ie $s$-wave and even in Matsubara frequency. 

\begin{figure}[tb]
    \centering
    \includegraphics[width=0.55\linewidth]{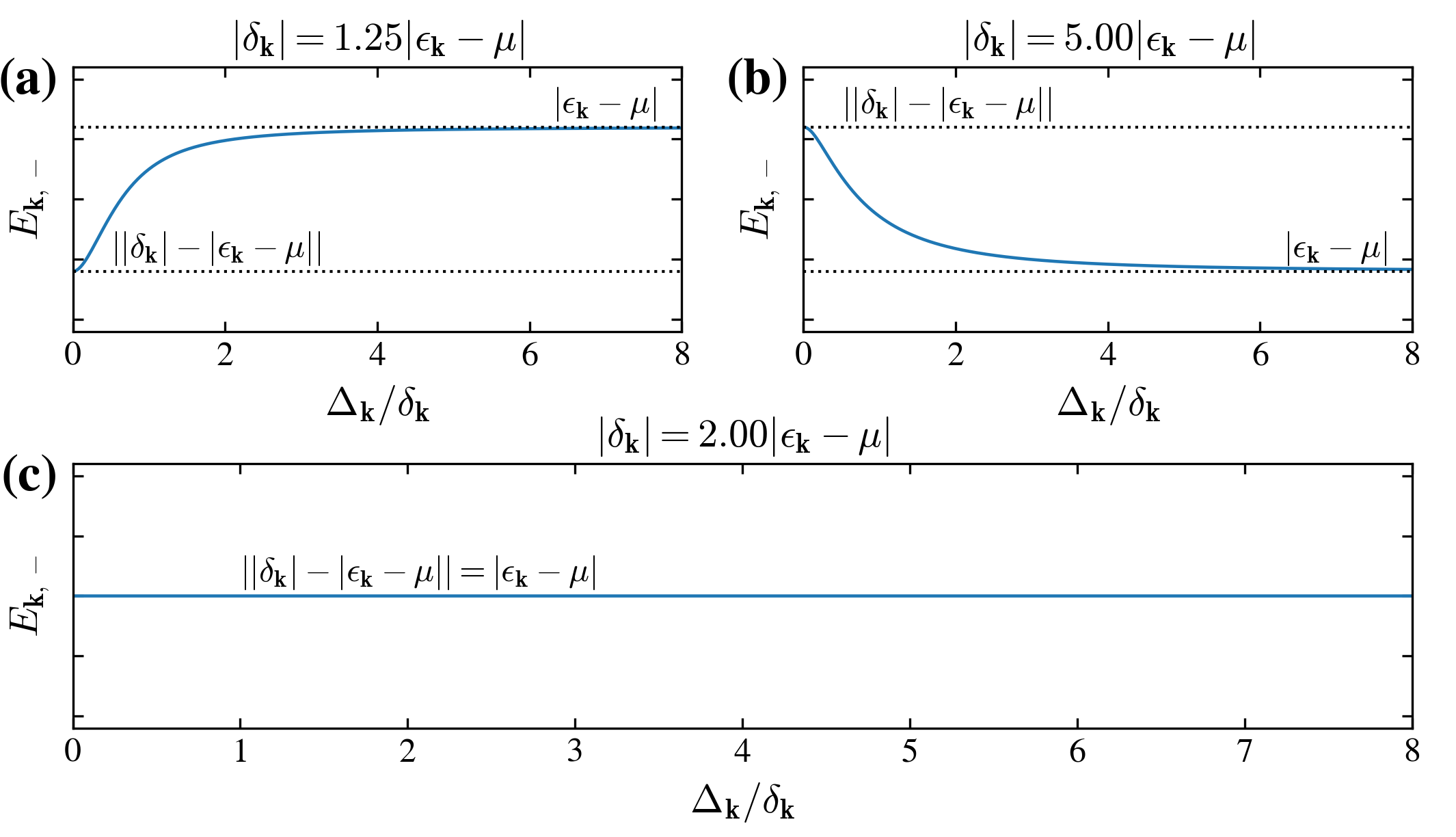}
    \caption{Plot of the particle-like minus band $E_{\vect{k},-}$ of the BdG spectrum calculated under the $\sigma_0+\sigma_x$ order parameter at fixed momentum as a function of $\Delta_{\vect{k}}$. The different panels show the increasing, decreasing and constant behavior caused by the different ratios of $\vert\delta_{\vect{k}}\vert$ and $\vert\epsilon_{\vect{k}}-\mu\vert$.}
    \label{fig:BdG_sig_0+sig_x}
\end{figure}

\begin{figure}[tb]
    \centering
    \includegraphics[width=0.5\linewidth]{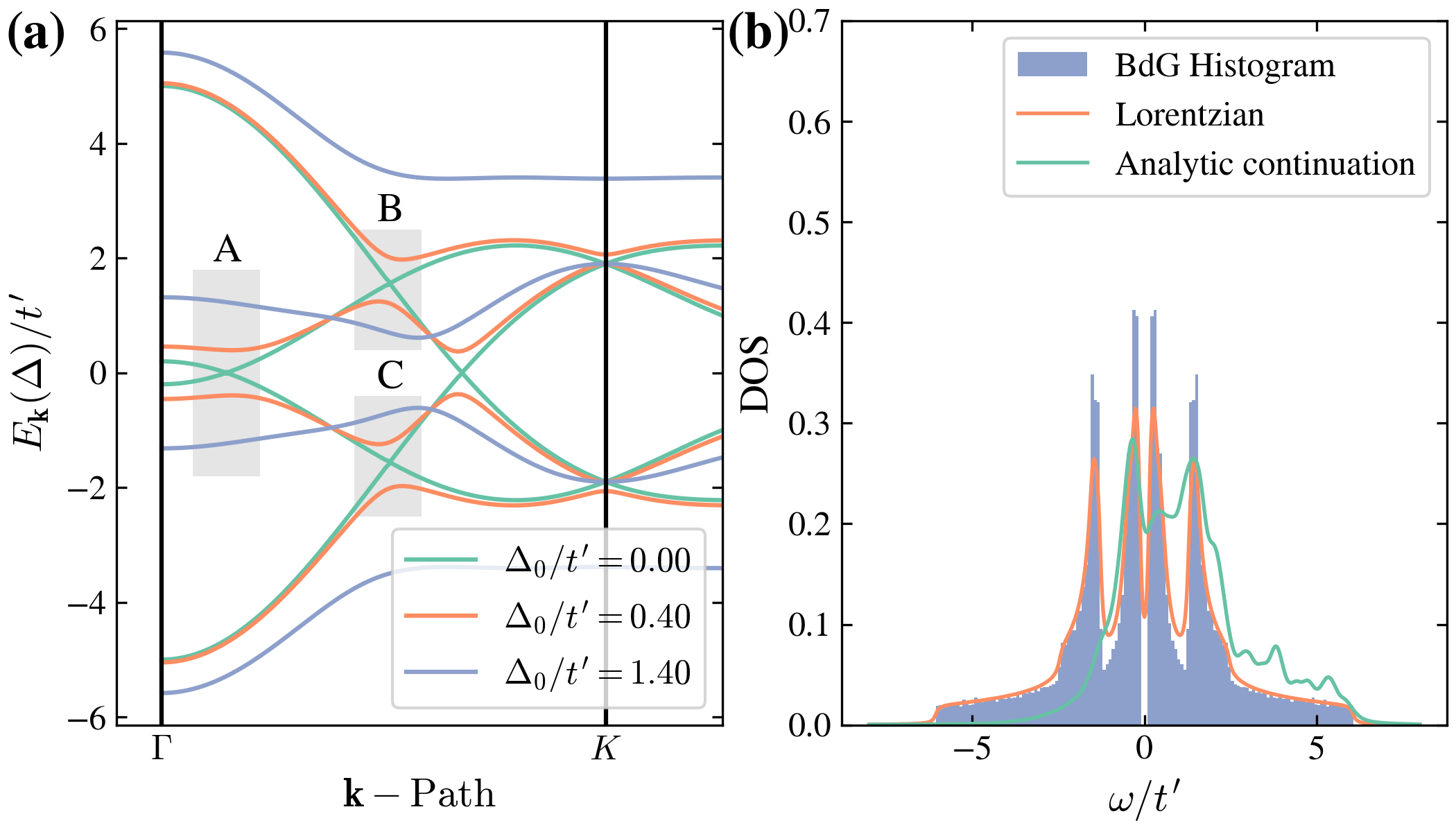}
    \caption{We show the four BdG bands (a) along the $\Gamma-K-M$ path for different values of the momentum independent order parameter $\Delta_{0}$. The shaded regions indicate avoided crossings caused by the diagonal (A) and off-diagonal (B,C) components of the order parameter. The bandsplitting and chemical potential were chosen such that the effects are well visible. Explicitly, we used $t/t'=0.8$ and $\mu/t'=0.4$. The DOS plot (b) was obtained using the BdG spectrum (blue and orange) as well as the spectral functions (green) obtained from the $\sigma_0+\sigma_x$ state at $T/T_c=0.95$ and $t/t'=0.61$.}
    \label{fig:BdG_k-path_sigx+sig0}
\end{figure}

A mean field analysis leads to the four BdG bands
\begin{equation}\label{eq:bdg_spec_sig0+sigx}
    \pm E_{\vk,\pm} = \pm\bigg[\delta_\vk^2 +(\epsilon_\vk-\mu)^2 + 2 \Delta_\vk^2 \pm 2\sqrt{\Delta_\vk^4+\delta_\vk^2\left(\Delta_\vk^2+(\epsilon_\vk-\mu)^2\right)}\bigg]^{1/2}.
\end{equation}
already showing some unusual behavior when when $2\vert\epsilon_\vk\vert \leq \vert\delta_\vk\vert$. In this regime the excitation gap at fixed momentum $\vk$ is decreasing for bigger order parameters. If the equality is true or $\delta_{\vk}=0$, the minus bands will be completely independent of $\Delta_{\vk}$ and the excitation gap is constant as a function of the order parameter. This can be seen by analyzing evaluating the small and big $\Delta_{\vect{k}}$ limits of the minus bands of \equref{eq:bdg_spec_sig0+sigx}:
\begin{equation}\label{eq:bdg_spec_sig0+sigx_bounds}
    E_{\vect{k},-}=
    \begin{cases}
        \big\vert\vert \delta_{\vk} \vert - \vert \epsilon_{\vect{k}}-\mu\vert\big\vert & \Delta_{\vect{k}} = 0 \\
        \vert \epsilon_{\vect{k}}-\mu\vert & \Delta_{\vect{k}} \rightarrow \infty
    \end{cases}.
\end{equation}
Further analysis shows that $E_{\vect{k},-}(\Delta_{\vect{k}})$ is always a monotonous function of $\vert\Delta_{\vect{k}}\vert$ and therefore the relation between the two limits dictates whether $E_{\vect{k},-}$ increases or decreases when dialing up the order parameter. 
We illustrate this behavior in \cref{fig:BdG_sig_0+sig_x}.
It can be explained by analyzing the avoided crossings in the BdG spectrum. Figure \ref{fig:BdG_k-path_sigx+sig0}(a) shows the four BdG bands for different values of the order parameter. To illustrate the effects we use a constant $\Delta_0$ for all momenta. The non-vanishing values in all band-components of the order parameter imply that a crossing between all electronic and hole like bands will be avoided. In the shaded area (A), we see that the $E_{\vk,-}$ bands intersect for $\Delta_0=0$. Dialing up the value of the order parameter leads to gapping-out those bands. This happens due to the $(-,-)$ component of the order parameter. Due to symmetry, particle- and hole-like bands originating from the same electronic band always meet at zero energy and the diagonal components of the order parameter therefore always shift the bands away from the Fermi-level creating a gapped spectrum.

This symmetry constraint does not hold for the intersection of particle- and hole-like bands originating from different electronic bands. We see this in the areas (B) and (C). The off-diagonal components of the order parameter prevent this intersection for $\Delta_0\neq0$. As the intersection appears away from the Fermi-energy, one of the bands gets shifted towards the Fermi-energy and therefore the excitation decreases with increasing $\Delta_0$. At the $K$ point, $\delta_{\vk=K}=0$ and the two original electronic points are degenerate. When dialing up the $\Delta_0$ we note that the effects cancel out and the two $\pm E_{\vk,-}$ bands sit at the same spot, independent of $\Delta_0$.

As both bounds in \equref{eq:bdg_spec_sig0+sigx_bounds} are generally positive, the dispersion can never show any zeros for $\Delta_{\vect{k}}\neq 0$. However if $\vert\epsilon_{\vect{k}}-\mu\vert$ is close to $\vert\delta_{\vect{k}}\vert$ for some momenta the excitation spectrum becomes nodal in the low temperature limit limit.

The resulting DOS at $T=0.95T_c$ is shown in \cref{fig:BdG_k-path_sigx+sig0}(b).
Even in this temperature regime the excitation gap is rather small. Comparison with \figref{fig:ac_sigy_0.95} (b) shows that the band structure of the order parameter seems to lead to an even smaller gap than that of the BOD order parameter. However there exists no temperature regime where the gap fully closes withing mean field theory.
Furthermore, we found that the BdG bands are not always consistent with the peaks of the spectral functions in some intermediate temperature regime for this specific order parameter. This could either indicate that the BdG bands do not provide an accurate description of the spectral function in this regime or a numerical artifact of the analytic continuation.

\section{Computing the correction terms of the superfluid stiffness}\label{app:stiffness_correction}
Equation \eqref{eq:stiffness} can be derived within linear response theory. However, as pointed out in \refcite{huhtinen_revisiting_2022}, it is crucial to carefully account for the dependence of all quantities on the vector potential. In this appendix, we address corrections arising from the dependence of the order parameter, $\Delta_\alpha(\vq)$, on the Cooper pair momentum $\vq$. 
We focus on the case where time reversal symmetry is present, which implies $\Delta_\alpha(\vq)=\Delta_\alpha^*(-\vq)$. Consequently, the total derivatives $\left.\frac{d \Delta_\alpha}{dq_i}\right|_{\vq=0} = i\left.\frac{d \Delta_\alpha^I}{dq_i}\right|_{\vq=0}$ are purely imaginary with $\Delta_\alpha^I$ denoting the imaginary part of the order parameter. In single-band models, this derivative can be eliminated by choosing a real-valued order parameter. However, in multi-band systems, such simplifications are generally not possible, as only one component of $\Delta_\alpha(\vq)$ can be made real at given $\vq$.
We set the global phase such that $\Delta_1(\vq) \in \mathbb{R}$ and utilize the fact that both components coincide at $\vq=0$, \ie $\Delta_1(0)=\Delta_2(0)=\Delta \in \mathbbm{R}$. The additive correction term to the superfluid stiffness, \equref{eq:stiffness}, can then be evaluated as
\begin{align} \label{eq:stiffness_corr}
    \delta D_s^{ij}= &
    \frac{1}{4} \sum_\vk\Bigg[ 
    -\frac{2 \, (d_i \imdt) (d_j \imdt) \, \sinh(\beta \sqrt{\xi_\vk^2 + \Delta^2})}{\sqrt{\xi_\vk^2 + \Delta^2} \left( \cosh(\beta \delta_\vk) + \cosh(\beta \sqrt{\xi_\vk^2 + \Delta^2}) \right)} \\ \nonumber
    -&\sum_\pm \tanh\left( \frac{\beta}{2} E_{\vk, \pm}  \right)  \left( 
    \frac{(d_i \imdt)(d_j \imdt)}{E_{\vk, \pm}} 
    + \frac{2  \left((\partial_i \theta_\vk)(d_j \imdt) + (d_i \imdt)(\partial_j \theta_\vk)\right)  \delta_\vk  \Delta}{\xi_\vk^2 + \Delta^2 \pm \delta_\vk \sqrt{\xi_\vk^2 + \Delta^2}} 
    \right) 
    \Bigg] \\
    -& \frac{2}{U} (d_i \imdt)(d_j \imdt)
\end{align}
where $d_i\imdt \equiv\left.\frac{d \imdt}{d\vq_i}\right|_{\vq=0}$. $U<0$ is the attractive on-site interaction. Within our momentum-space model, there is no canonical choice for this parameter. However, as mentioned in the main text the corrections arising from \equref{eq:stiffness_corr} are negligible compared to the main contribution stated in \equref{eq:stiffness}.

As expected the term only depends on the derivatives of the imaginary part of the second component of the order parameter, \ie $\imdt$. These derivatives are highly nontrivial to compute, as they seems to require solving the gap equation at non-zero Cooper pair momentum $\vq$. Here, we neglect the explicit electron-momentum dependence $\vk$.

We start the discussion by studying the BdG Hamiltonian \eqref{eq:H_BdG_maintext} at finite $\vq$
\begin{equation}\label{eq:HBdG_q}
    H_\mathrm{BdG}(\vq) = \begin{pmatrix}
        H^+_{\vq + \vk} & \hat{\Delta}\\
        \hat{\Delta}^\dagger & -\left(H^-_{\vq-\vk}\right)^*
    \end{pmatrix}
\end{equation}
with valley Hamiltonian $H^\pm_\vk = \sigma_0 \xi_\vk \pm(\sigma_1 \delta^1_\vk +\sigma_2 \delta^2_\vk)$ and band diagonal order parameter $\hat{\Delta} = \mathrm{diag(\Delta_1, \Delta_2)}$. Following \refcite{huhtinen_revisiting_2022} the desired derivative can be rewritten as
\begin{equation}\label{eq:dDelta_dq}
    \frac{d \Delta_2^I}{dq_i} = -\frac{\partial^2 \Omega}{\partial q_i \partial \imdt} / \frac{\partial^2\Omega}{\partial\imdt^2}
\end{equation}
which involves only partial derivatives of the grand canonical potential 
\begin{equation}
    \Omega=  -\frac{1}{\beta} \sum_\vk \sum_i \ln \left[1+\exp \left(-\beta E_{\vk, i}\right)\right]
    +\sum_\vk \operatorname{Tr} H_\vk^{-}-n N_c \mu-N_c \sum_\alpha \frac{\left|\Delta_\alpha\right|^2}{U}.
\end{equation}
The $\vq$ dependence is hidden in den eigenvalues $E_{\vk,i}$ of \equref{eq:HBdG_q}.
Taking the respective derivatives yields the terms
\begin{equation}\label{eq:d2Omega_dqdDelta}
    \frac{\partial^2 \Omega}{\partial q_i \partial \imdt} = \sum_\vk\sum_j 
    \frac{\partial_{\imdt}\partial_{q_i}E_{\vk, j} + e^{\beta E_{\vk,j}}\left[\partial_{\imdt}\partial_{q_i}E_{\vk, j} - \beta(\partial_{\imdt}E_{\vk, j})(\partial_{q_i}E_{\vk, j})\right]}
    {(1+e^{\beta E_{\vk,i}})^2}
\end{equation}
and

\begin{equation}\label{eq:d2Omega_dDelta2}
    \frac{\partial^2 \Omega}{\partial \imdt^2} = \sum_\vk\sum_j 
    \frac{\partial_{\imdt}^2 E_{\vk, j} + e^{\beta E_{\vk,j}}\left[\partial_{\imdt}^2 E_{\vk, j} - \beta(\partial_{\imdt}E_{\vk, j})^2\right]}
    {(1+e^{\beta E_{\vk,i}})^2} - \frac{2 N_c}{U}.
\end{equation}
The derivatives of the eigenvalues require some special attention as it is generally not possible to analytically diagonalize the general Hamiltonian \eqref{eq:HBdG_q} at finite $\vq$ and $\Delta_1\neq\Delta_2$. However, as all terms are eventually evaluated at $\vq=0$ and $\Delta_1=\Delta_2=\Delta \in \mathbbm{R}$ it is still possible to find exact analytical expressions by perturbative methods. We start by rewriting \equref{eq:HBdG_q} as
\begin{equation}
    H_\bdg(\vq, \imdt) = H_0 + \imdt H_1, \qquad \mathrm{with}\quad H_0=H_\bdg(\vq=0)\vert_{\imdt=0}, \, H_1=
    \begin{pmatrix}
        0 & 0 & 0 & 0\\
        0 & 0 & 0 & i\\
        0 & 0 & 0 & 0\\
        0 & -i & 0& 0\\
    \end{pmatrix}.
\end{equation}
The eigensystem of $H_0$ can be computed analytically and is denoted by $\left(E^0_i, \vert\psi^0_i\rangle\right)$. 
The first order corrections to the states read as
\begin{equation}
    \vert\psi^1_i\rangle = \Delta_2^I\sum_{j\neq i}\frac{\langle\psi_j^0\vert H_1 \vert\psi_i^0\rangle}{E^0_i-E^0_j}\vert\psi_j^0\rangle,
\end{equation}
and the full eigensystem is given by $\left(E_i, \vert\psi_i\rangle\right)$. For notational convenience we omit the $\vk$ index here. 

The easiest terms to evaluate are the first-order derivatives $\partial_\imdt E_j$ and $\partial_{q_i}E_j$, which can be computed using the general property, $\partial_x \lambda(x) = \langle\lambda(x) | \left[\partial_xM(x)\right]|\lambda(x)\rangle$, for a parameter-dependent matrix $M(x)$ and respective eigensystem $(\lambda(x),\vert\lambda(x)\rangle)$. Applying this we find
\begin{align}
    \label{eq:dqE}
    \partial_{q_i} E_j\vert_{\vq=0,\imdt=0} =& \langle \psi_j^0\vert (\partial_{q_i}   H_\mathrm{BdG}\vert_{\vq=0})\vert \psi_j^{0}\rangle\\
    \label{eq:dDeltaE}
    \partial_{\imdt} E_j\vert_{\vq=0,\imdt=0} =& \langle \psi_j^0\vert H_1\vert \psi_j^{0}\rangle.
\end{align}

To compute the second-order derivatives, we must account for the dependence of the states on $\imdt$. Specifically, the terms of interest are of the form $\partial_{\imdt}|\psi_i\rangle\vert_{\imdt=0}$, which means that only the terms linear in $\imdt$ contribute. Consequently, we can exactly rewrite $\partial_{\imdt}|\psi_i\rangle\vert_{\imdt=0} = \frac{1}{\imdt} |\psi_i^1\rangle$. The remaining derivatives are therefore found as
\begin{align}
    \label{eq:dDeltadqE}
    \partial_\imdt\partial_{q_i} E_j\vert_{\vq=0,\imdt=0} &= \langle \psi_j^1\vert (\partial_{q_i}   H_\mathrm{BdG}\vert_{\vq=0})\vert \psi_j^{0}\rangle + \langle \psi_j^0\vert (\partial_{q_i}   H_\mathrm{BdG}\vert_{\vq=0})\vert \psi_j^{1}\rangle \\
    \label{eq:dDelta2E}
    \partial_{\imdt}^2 E_j\vert_{\vq=0,\imdt=0} &= \langle \psi_j^1\vert H_1\vert \psi_j^{0}\rangle + \langle \psi_j^0\vert H_1\vert \psi_j^{1}\rangle.
\end{align}
With the equations (\ref{eq:dqE}, \ref{eq:dDeltaE}, \ref{eq:dDeltadqE}, \ref{eq:dDelta2E}) we find exact expressions for all terms in \equref{eq:d2Omega_dqdDelta} and \equref{eq:d2Omega_dDelta2} and thus $\frac{d \imdt}{dq_i}$ in \equref{eq:dDelta_dq} can be computed without relying on numerical differentiation and without solving the gap equation at nonzero $\vq$.

\section{Low temperature scaling of superfluid stiffness in the chiral state}\label{sec:app_low_temp_scaling}
In this appendix we discuss the low temperature scaling behavior of the superfluid stiffness \eqref{eq:stiffness} in the chiral state. We split the expression as

\begin{align}
    \label{eq:D_conv1}
    D_{s}^\conv{1} &= \sum_{\vk}\frac{\lambda_\vk^\mathrm{conv}}{\sqrt{\xi_\vk^2+|\Delta_\vk|^2}} \sum_\pm \tanh\left(\frac{\beta}{2}E_{\vk,\pm}\right)\\
    \label{eq:D_conv2}
    D_{s}^{\mathrm{conv_2}} &= -\frac{\beta}{2}\sum_{\vk}\lambda_\vk^\mathrm{conv} \sum_\pm \frac{1}{\cosh ^2\left[\frac{\beta}{2}E_{\vk, \pm}\right]} \\
    \label{eq:D_geo1}
    D_{s}^{\mathrm{geo_1}} &= \sum_{\vk}\lambda_\vk^\mathrm{geo}\delta_\vk\sum_\pm \pm \tanh\left(\frac{\beta}{2}E_{\vk,\pm}\right)\\
    \label{eq:D_conv2}
    D_{s}^{\mathrm{geo_2}} &=-\sum_{\vk}\lambda_\vk^\mathrm{geo}\delta_\vk^2\sum_{\pm}\frac{\tanh \left[\frac{\beta}{2}E_{\vk, \pm}\right]}{E_{\vk, \pm}}
\end{align}
where the superscript discriminates between the two terms of the conventional and geometric contribution respectively. The prefactors have been absorbed into
\begin{align}
    \lambda_\vk^\mathrm{conv} &= \left[\left(\partial_i\xi_\vk\right)\left(\partial_j \xi_\vk\right)-\left(\partial_i \delta_\vk\right)\left(\partial_j \delta_\vk\right)\right] \frac{|\Delta_\vk|^2}{\xi_\vk^2+|\Delta_\vk|^2}\\
    \lambda_\vk^\mathrm{geo} &=\left(\partial_i \theta_\vk\right)\left(\partial_j \theta_\vk\right) \frac{|\Delta_\vk|^2}{\xi_\vk^2+|\Delta_\vk|^2}
\end{align}
The terms proportional to the derivative of the order parameter are given by
\begin{align}
    \label{eq:D_Delta1}
    D_s^{\Delta_1} &= -\sum_\vk\lambda^\Delta_\vk\frac{\xi_\vk^2}{\sqrt{\xi_\vk^2 + |\Delta_\vk|^2}}\sum_\pm\tanh\left(\frac{\beta}{2}E_{\vk, \pm}\right) \\
    \label{eq:D_Delta2}
    D_s^{\Delta_2} &= -\sum_\vk\lambda^\Delta_\vk|\Delta_\vk|^2\frac{\beta}{2}\sum_\pm\frac{1}{\cosh^2{\left(\frac{\beta}{2}E_{\vk, \pm}\right)}}\\
    \label{eq:D_phi}
    D_s^\varphi &= -\sum_\vk\lambda^\varphi_\vk\sum_\pm\tanh\left(\frac{\beta}{2}E_{\vk, \pm}\right).
\end{align}
Besides the prefactors,
 \begin{align}
     \lambda_\vk^\Delta &= \left(\partial_i |\Delta_\vk|\right)\left(\partial_j |\Delta_\vk|\right)\frac{1}{\sqrt{\xi^2 + |\Delta_\vk|^2}}\\
     \lambda_\vk^\varphi &=\left(\partial_i \varphi_\vk\right)\left(\partial_j \varphi_\vk\right) \frac{|\Delta_\vk|^2}{\sqrt{\xi_\vk^2+|\Delta_\vk|^2}} ,
 \end{align}
\equref{eq:D_Delta1}, \equref{eq:D_phi} and \equref{eq:D_Delta2} are equivalent to the conventional contributions \equref{eq:D_conv1} and \equref{eq:D_conv2}, respectively. Consequently those terms do not yield any new temperature dependence and could be absorbed in the respective conventional terms by redefining $\lambda_\vk^\conv{}$. However, to stay consistent with only including dispersion-derivatives in the conventional contribution, we will, without loss of generality, keep the terms separate and omit $D_s^{\Delta_{1/2}}$ and $D_s^\varphi$ from the following discussion.

For simplicity we assume isotropic dispersions which implies that $D_s^{ij}\propto \delta_{ij}$. 
In the regime where both, $E_{\vk, +}$ and $E_{\vk, -}$ are gapped, it holds that $\beta E_{\vk, \pm} \gg 1 \,, \forall \vk \in \mathrm{BZ}$ at small temperatures. As a consequence all deviations $D_s(T)-D_s(0)$ are exponentially suppressed. However in the nodal regime, near the Fermi surface this does not hold anymore, which might lead to non exponential scaling.
To understand how the terms behave, we model $E_{\vk, -} = v_F (k-k_F)$ with $k=\vert\vk\vert$ and Fermi velocity and momentum $v_F>0$ and $k_F>0$, respectively. As long as the terms decay rapidly as a function of $|\vk|$, we can approximate $\sum_\vk\rightarrow \int_0^{2\pi}d\phi\int_0^\infty dk k =2\pi\int_0^\infty dk k$.
We start with the first conventional contribution \ie $D_s^\conv{1}$. We rewrite the term
\begin{equation*}
    \sum_\pm\tanh\left(\frac{\beta}{2}E_{\vk,\pm}\right) = \frac{1}{1+e^{-\beta E_{\vk,-}}} + \mathcal{O}(e^{-E_0\beta})
\end{equation*}
where $E_0$ represents the minimum of band energy scales. For simplicity we assume $E_0>0$. In the zero temperature limit this terms becomes the Heaviside step function $\Theta(E_{\vk,-}) =\Theta(k-k_F)$. We can therefore model the low temperate deviations as
$D_s^\conv{1}(T)-D_s^\conv{1}(0) =2\pi\int_0^\infty dk k \lambda_k^\conv{} \left( \Theta(k-k_f) - \frac{1}{1+e^{-\beta v_F (k-k_F)}}\right) + \mathcal{O}(e^{-E_0\beta})$. Substituting $x=\beta v_F (k-k_F)$ and using $\lambda_k^\conv{}=\lambda_{-k}^\conv{}$ we find that the leading contribution scales as $T^2$. The same argument holds for $D_s^\geo{1}$, which only differs by a relative minus sign between the two summands. As a consequence we find quadratic low temperature scaling with opposite sign than $D_s^\conv{1}$.

The second conventional contribution, $D_s^\conv{2}$, consists of two summands. However as $E_{\vk, +}>0$ holds by construction only the term corresponding to $E_{\vk,-}$ potentially features non exponential behavior. Due to the hyperbolic cosine term the integrand is localized in a small window around the Fermi surface which justifies rewriting the sum as $D_s^\conv{2}(T)= 2\pi\int_{-\infty}^\infty dk k \lambda_k^\conv{} \frac{\beta}{2}\frac{1}{\cosh^2(\beta v_F (k-k_F)/2)} + \mathcal{O}(e^{-E_0\beta})$, where we extended the lower bound from $0\rightarrow-\infty$. We can elegantly understand this term by recognizing $\lim_{a\rightarrow \infty} \frac{a}{2}\mathrm{sech}^2(ax)=\delta(x)$ is a representation of the Dirac distribution $\delta(x)$. Substituting $x=\frac{v_F}{2}k$ and defining $\widetilde{\lambda}_x=\lambda^\conv{}_{k=2x/v_F} $ we can interpret the integral, up to a constant prefactor, as a convolution $(x\widetilde{\lambda}_x * f_\beta)(x_F)$, where $f_\beta(x) = \frac{\beta}{2} \mathrm{sech}^2(\beta x)$ and $x_F = \frac{v_F}{2}k_F$. The error resulting from big but finite $\beta$ is given by the moments of the distribution $f_\beta(x)$. Due to the symmetric nature of $f_\beta(x)$ all odd moments vanish and the first non-zero contribution is generically given by the second moment which scales as $\frac{1}{\beta^2}$. We therefore find $(x \widetilde{\lambda}_x*f_\beta) (x_F) = x_F\widetilde{\lambda}_{x_F} + c_2 T^2 +c_4 T^4  + \dots$, $c_2, c_4\in \mathbbm{R}$. The first term is the result one would obtain if $f_\beta(x)$ was a sharp $\delta(x)$, \ie the zero temperature limit. The prefactor $c_n$ is generically non-zero if and only if $x\widetilde{\lambda}_x$ includes terms of at least $x^n$. In order to capture non-exponential behavior in $D_s^\conv{2}(T)-D_s^\conv{2}(0)$, $\widetilde{\lambda}_x$ must therefore include at least linear terms in $x$ (due to $\widetilde{\lambda}_x=\widetilde{\lambda}_{-x}$ at least $x^2$ in our model). The latter follows from $\lambda_k^\conv{}=\lambda_{-k}^\conv{}$.

In the low temperature limit the integrands of $D_s^\geo{2}$ do not decay rapidly and the integral is UV divergent. Thus it is necessary to introduce a momentum cutoff $\Lambda$, which enters as $D_s^\geo{2}(T)\sim \log(\Lambda v_F / T)$ for small temperatures. This leads to non-universal behavior and the scaling depends on the dispersions far away from the BdG Fermi surface. We leave the analysis of the consequences for the superconductor at and below the associated exponentially small temperatures for future work since at such low energy scales various other perturbations, such as inhomogeneities, are likely more relevant.

Summing up the analysis we find the universal scaling
\begin{align}
    &D_{s}^\conv{1} (T)-D_{s}^{\mathrm{conv_1}}(0) \propto -T^2 \label{eq:Dconv1}\\
    &D_{s}^{\mathrm{geo_1}}(T)-D_{s}^{\mathrm{geo_1}}(0) \propto T^2\label{eq:Dgeo1}\\
    &D_{s}^{\mathrm{conv_2}}(T)-D_{s}^{\mathrm{conv_2}}(0) \propto 
    \begin{cases}
        T^2 & \text{if $\lambda_k^\conv{}$ includes $k^1$ or higher orders}\\
        \mathcal{O}\left(e^{-\frac{E_0} {T}}\right) & \mathrm{else}
    \end{cases}\label{eq:Dconv2}
\end{align}
The signs of \eqref{eq:Dconv1} and \eqref{eq:Dgeo1} could be swapped for different system parameters. However the relative negative sign between the terms persists. As emphasized above, $D_s^{\Delta_{1/2}}$ features the same scaling behavior as $D_s^{\conv{1/2}}$.$D_s^\varphi$ scales like $D_s^\conv{1}$.
\begin{figure}
    \centering
    \includegraphics[width=0.5\linewidth]{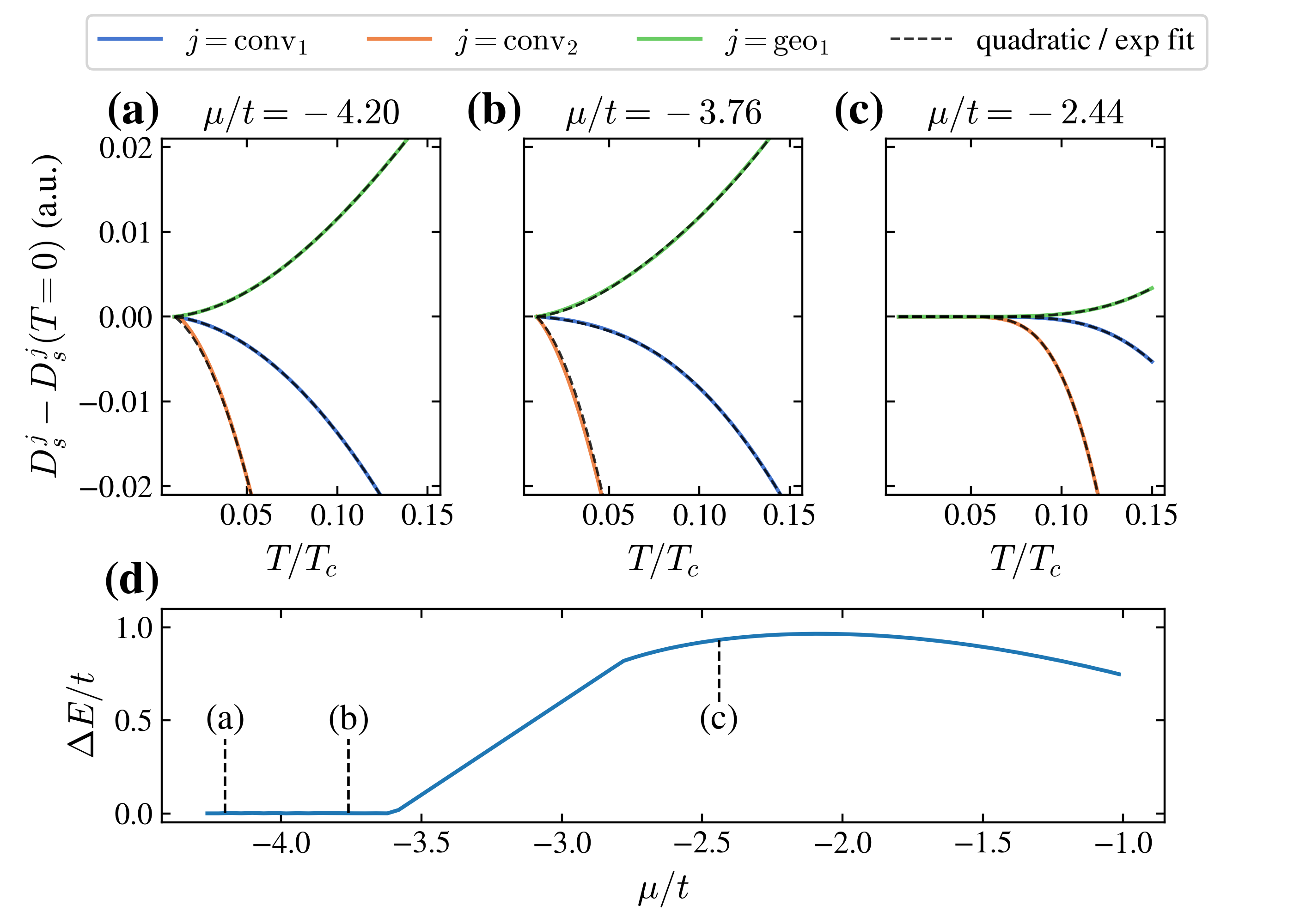}
    \caption{Terms of the superfluid stiffness that feature universal scaling in the nodal regime (panel (a)), close to the boundary (panel (b)) and in the gapped regime (panel (c)) of the chiral, BOD state. The dashed lines are obtained by fitting the results of the scaling analysis in in the respective regimes, \ie $y=u\tau^2 + ve^{-w/\tau}$ in panels (a) and (b) and $y=ue^{-v/\tau}$ in panel (c), where we defined the dimensionless temperature $\tau \equiv T/T_c$. Panel (d) shows the excitation gap, $\Delta E \equiv \min_{\vk\in\mathrm{BZ}}\left[|\delta_\vk - \sqrt{\xi_\vk^2 +|\Delta_\vk|^2}|\right]$, as a function of the chemical potential.}
    \label{fig:app_stiffness_scaling}
\end{figure}
\figref{fig:app_stiffness_scaling} (a)-(c) shows the terms (\ref{eq:Dconv1}-\ref{eq:Dconv2}) plotted separately. To compare the scaling behavior we subtract the (approximate) zero temperature values of all terms individually \ie plot $D^j_s(T)-D^j_s(T_0)$ where $T_0=\frac{1}{80}T_c$ is the lowest computed temperature value, $T_c/t=2.5$.
Panel (d) shows the excitation gap as a function of $\mu$. In panel (a), the system is well inside the nodal regime, the solid lines show the scaling of $D_s^\conv{1}$, $D_s^\geo{1}$ and $D_s^\conv{2}$, respectively. The dashed lines are given by $y=u\tau^2 + v e^{-w/\tau}$ where $u, v,w$ are fit parameters and $\tau \equiv T/T_c$. The numerical values, $(u^{(a)}_\conv{1}, v^{(a)}_\conv{1}, w^{(a)}_\conv{1}) \approx (-8.78,  0.00,  0.12)$ show, that the exponential corrections to $D_s^\conv{1}$ are small and the curve is dominated by the quadratic term in the entire plotted temperature regime. Near the transition to the gapped regime the minimal finite energy scale $E_0$ is naturally smaller and the exponential terms $e^{-\beta E_0}$ start contributing sooner. We show this in panel (b) which optically looks similar to (a), however the the fit parameters, $(u^{(b)}_\conv{1}, v^{(b)}_\conv{1}, w^{(b)}_\conv{1}) \approx (-4.45, -0.17,  0.19)$ show, that $D_s^\conv{1}$ does not scale as a pure parabola on the entire plotted temperature scale. 
For $D_s^\geo{1}$ we find the fit parameters $(u^{(a)}_\geo{1}, v^{(a)}_\geo{1}, w^{(a)}_\geo{1}) \approx (7.62, -0.12,  0.21)$ and $(u^{(b)}_\geo{1}, v^{(b)}_\geo{1}, w^{(b)}_\geo{1}) \approx(8.74, -0.08,0.15)$ in the regimes (a) and (b), respectively. We note that the sign of the quadratic term is, as expected, opposite to the sign of the first conventional contribution.
From the scaling analysis alone, it is not clear how the deviations of $D_s^\conv{2}$ behave at low temperatures, as it depends on the form of $\lambda_k^\mathrm{conv}$. Panels (a) and (b) show a fit with $(u^{(a)}_\conv{2}, v^{(a)}_\conv{2}, w^{(a)}_\conv{2}) \approx (-49.35, 1.28, 0.25)$ and $(u^{(b)}_\conv{2}, v^{(b)}_\conv{2}, w^{(b)}_\conv{2}) \approx (-63.54, 0.49, 0.15)$ to the curve, assuming that $\lambda_k^\conv{}$ includes $k^2$ or higher order terms. 
Visually, the curves coincide, however without analyzing $\lambda_k^\conv{}$ one cannot be sure that the correct function was optimized. However as this analysis is concerned with the universal scaling properties and $D_s^\conv{2}$ does not yield new contributions in both cases, we shall not analyze this further.
Panel (c) shows the terms well inside the gapped regime. All curves clearly saturate. The red lines shows a fit $y = u e^{-v/\tau}$ of $D_s^\conv{1,2}$ and $D_s^\geo{1}$ with $(u^{(c)}_\conv{1}, v^{(c)}_\conv{1}) \approx(-1.00,  0.32)$, $(u^{(c)}_\conv{2}, v^{(c)}_\conv{2}) \approx(-5.03,  0.26)$ and $(u^{(c)}_\geo{1}, v^{(c)}_\geo{1}) \approx (0.41, 0.29)$.

\end{appendix}

\end{document}